\documentclass[a4paper]{article}

\usepackage{amsfonts,amsmath,amssymb}
\usepackage{graphicx}
\usepackage{hyperref}
\usepackage[font={small}]{caption}
\usepackage{epstopdf} %only when using miktex
\usepackage{color}

\def\beq{\begin{equation}}
\def\eeq{\end{equation}}
\def\Rmax{R_{\mathrm{max}}}
\def\Rmin{R_{\mathrm{min}}}
\def\D{\mathrm d}
\def\E{\mathrm e}

\title{Pseudo-spectral construction of non-uniform black string solutions in five and six spacetime dimensions}
\author{Michael Kalisch and Marcus Ansorg \\ \ \\
		\textit{\small{Theoretisch-Physikalisches Institut, Friedrich-Schiller-Universit\" at Jena,}} \\ 
		\textit{\small{Max-Wien-Platz 1, D-07743 Jena, Germany}} \\
		\textit{\small{E-mail: michael.kalisch@uni-jena.de, marcus.ansorg@uni-jena.de}}}
\date{}

\begin{document}

\maketitle

\begin{abstract}
	In this paper, we describe in detail a scheme for the construction of highly accurate numerical solutions to Einstein's field equations in five and six spacetime dimensions corresponding to non-uniform black strings. The scheme consists of a sophistically adapted multi-domain pseudo-spectral method which incorporates a detailed understanding of the solution's behavior at the domain boundaries and at critical points. In particular, the five-dimensional case is exceedingly demanding as logarithmic terms appear which need to be treated with special care. The scheme resolves these issues and permits the investigation of unprecedentedly strong deformations of the black string horizon. As a consequence, we are able to study in detail the critical regime in phase diagrams displaying characteristic   thermodynamic quantities such as mass and entropy. Our results show typical spiral curves in such diagrams which provides a strong support of previous numerical works.
\end{abstract}

%%%%%%%%%%%%%%%%% 
\newpage
\tableofcontents

\section{Introduction}
\label{sec:Introduction}

The major motivation for the construction of stationary black hole solutions in higher spacetime dimensions $D>4$ originates from considerations in string theory and the AdS/CFT correspondence. While analytic solutions are rare in this context, much effort has been expended to solve Einstein's equations on the computer, see the recent review~\cite{Dias:2015nua}. 

Naturally, numerical calculations are accompanied with errors, thus leading inescapably to a mere approximation of the true solution to the problem in question. Due to finite computational resources, the error can not always be pushed below a reasonable value. In particular, when the underlying mathematical structure is more involved (e.g.,~if a strongly pronounced peak appears), then an accurate numerical solution becomes more challenging and at some critical point the method may fail. Unfortunately, this typically happens when specifically interesting branch points or phase transitions are encountered. Consequently, in a situation where standard algorithms reach their limitations, it arises the necessity to design a method which is specifically adapted to the mathematical circumstances. With the help of a corresponding code developed along these lines, the critical regime can be explored and unrevealed properties may become manifest.

A situation of this kind arises in the context of non-uniform black strings (henceforth ``NBS'') in the limit of maximal deformation of the horizon. An NBS emanates from the Gregory-Laflamme (henceforth ``GL'') instability~\cite{Gregory:1993vy} of a uniform black string (henceforth ``UBS''), which can be described as a product of a Schwarzschild-Tangherlini solution with a circle. This instability occurs when the mass of the string is sufficiently small compared to the size of the circle and it leads to a deformation of the string's horizon along the compact dimension, see~\cite{Harmark:2007md,Gregory:2011kh} for reviews. 

Static NBS solutions were first obtained approximatively with the help of perturbation theory in five spacetime dimensions $D=5$. In the corresponding article \cite{Gubser:2001ac} an appropriate measure of non-uniformity was introduced: 
\beq
	\lambda = \frac{1}{2} \left( \frac{R_\text{max}}{R_\text{min}} -1 \right) \, .
	\label{eq:lambda}
\eeq  
Here, $R_\text{max}$ and $R_\text{min}$ present maximal and minimal radii of the black string along the compact dimension.  Naturally, UBS are characterised by $R_\text{max}=R_\text{min}$, i.e.~$\lambda = 0$, while for NBS solutions we have $\lambda >0$.

The procedure described in~\cite{Gubser:2001ac} was later applied to higher dimensions~\cite{Wiseman:2002zc,Sorkin:2004qq}. Going beyond the scope of perturbation theory which considers small horizon deformations ($\lambda\ll 1$), numerical constructions of NBS in different spacetime dimensions were performed in a number of works \cite{Wiseman:2002zc,Kleihaus:2006ee,Sorkin:2006wp,Headrick:2009pv,Figueras:2012xj,Kalisch:2015via}. Interestingly, the results suggest that along the NBS branch the ratio $R_\text{min}/R_\text{max}$ shrinks gradually down to zero, i.e.~NBS solutions exist for all $\lambda\in [0,\infty)$. In the limit $\lambda\to\infty$ a curvature singularity is encountered at which the horizon pinches off, and this is exactly the region described above where the numerics reaches its limitations. On the other hand, this limit is of particular interest 
as it describes, according to a conjecture presented in \cite{Kol:2002xz}, a phase transition to another branch of solutions, called  localized or caged black hole branch. This designation arises from the fact that the horizons of the corresponding objects do not extend to the entire compact dimension, see~\cite{Kol:2003if}. Numerical results provide evidence in favor of this phase  transition~\cite{Wiseman:2002ti,Sorkin:2003ka,Kudoh:2003ki,Kudoh:2004hs,Headrick:2009pv}. Moreover, the local geometry of the transit solution at the pinch-off point is conjectured to be that of a double-cone~\cite{Kol:2002xz} (see~\cite{Asnin:2006ip} for further plausible arguments). Additional numerical evidence supporting this conjecture was presented in~\cite{Kol:2003ja,Sorkin:2006wp,Figueras:2012xj}. Very readable reviews regarding the black hole/black string phase transition can be found in \cite{Kol:2004ww,Harmark:2005pp,Horowitz:2011cq}.

In this paper we concentrate on the NBS branch. Motivated by contradictory results of previous works we investigate in particular the critical regime of strong horizon deformations $\lambda\gg 1$. To be more precise, the discrepancy concerns the question whether in five and six spacetime dimensions the mass in the NBS phase reaches a maximum for some finite $\lambda$ \cite{Kleihaus:2006ee,Sorkin:2006wp,Figueras:2012xj}. Our results show agreement for both dimensions with the findings reported in  \cite{Kleihaus:2006ee}, i.e.~we detect a clearly pronounced maximum. Moreover, we are able to identify two further turning points in the mass curve if $\lambda$ is further increased. Since other thermodynamic quantities show a similar behavior, we observe the beginning of a spiral curve in the black string phase diagram. 

In the case $D=6$, the aforementioned results have been outlined in~\cite{Kalisch:2015via}. Here we provide an elaborated description of the numerical algorithm and include the case $D=5$ which requires a specific treatment (to be depicted in Sec.~\ref{subsec:5D_implementation}) in order to deal with the logarithmic behavior of the solution. 

Our numerical scheme is based on pseudo-spectral methods and includes a number of sophisticated adaptations in order to resolve the critical regime satisfactorily. The techniques encompass several appropriate coordinate mappings (which provide, in particular, a compactification of infinity), the introduction of multiple domains and the split of each metric function into two parts (near infinity). In addition, we need high resolution near the critical spatial point at which the pinch-off occurs for $\lambda\to\infty$. We emphasize that the scheme is highly specialized to the construction of strongly deformed NBS. In principle, NBS solutions for $D>6$ should be constructible in the same manner as described here (with some slight modifications). Some of the individual tricks, however, may also be applicable to other problems. 

The paper is organized as follows. In section~\ref{sec:Physical_setup} we provide the form of the metric together with the system of equations to be solved in the sequel. Furthermore, for later convenience we quote relevant formulas and relations involving specific geometric and thermodynamic quantities. The numerical scheme is presented in section~\ref{sec:Numerical_implementation}. In particular, differences in the construction of five and six dimensional NBS are pointed out. The results are reported in Section~\ref{sec:Results}, including plots of characteristic geometric and thermodynamic variables. Also, we provide a discussion of the accuracies obtained. Finally, we conclude this article in Section \ref{sec:Conclusions}. More details on perturbation theory (see Sec.~\ref{appendix:sec:Perturbations_around_the_UBS}) and the numerical scheme (see Sec.~\ref{appendix:sec:Resolving_the_critical_point}) are provided in the appendix.

\section{Physical setup}
\label{sec:Physical_setup}

We consider the static NBS metric in $D$ dimensions and with the background $\mathbb R^{D-2,1}\times \mathbb S^1 $ in the form 
\beq
	\D s^2 = -\E ^{2A(r,z)}f(r)\D t^2 + \E ^{2B(r,z)} \left( \frac{\D r^2}{f(r)} + \D z^2 \right) + r^2\E ^{2C(r,z)}\D \Omega ^2_{D-3} \, , 
	\label{eq:line_element}
\eeq
where $\D \Omega ^2_{D-3}$ is the line element of a unit $(D-3)$-sphere. The three unknown metric functions $A$, $B$ and $C$ depend on the radial coordinate $r\in [r_0,\infty )$ and the periodic coordinate $z\in [0,L]$ varying along $\mathbb S^1$. With 
\beq
	f(r)=1-\left( \frac{r_0}{r} \right) ^{D-4}
	\label{eq:fct_f}
\eeq
the horizon resides at $r=r_0$. 

It becomes apparent that  $A\equiv B\equiv C\equiv 0$ corresponds to  the UBS metric (which can be seen as a product of a Schwarzschild-Tangherlini solution in $D-1$ dimensions with a circle of size $L$). Holding $r_0$ fixed, the UBS is subject to the GL instability if $L$ is larger than a critical value $L_{\text{GL}}$, though it is stable if $L<L_{\text{GL}}$. This instability breaks the translation invariance along the $z$-direction, thus leading to the NBS branch which can be described by non-vanishing potentials $A$, $B$ and $C$. 

\subsection{Field equations}
\label{subsec:Field_equations}

The NBS are described by specific vacuum solutions to Einstein's field equations. The Einstein tensor's components $G^\mu_\nu$ satisfy $G^t_t=0$, $G^r_r+G^z_z=0$ and $G^\theta _\theta=0$ ($\theta$ is an angle of the ($D-3$)-sphere) from which we get the following system of equations~\cite{Kleihaus:2006ee} (we use the notations $':= \partial /\partial r$ and $\dot{} := \partial /\partial z$):
\refstepcounter{equation} \label{eq:field_eqns}
\begin{align}
	0 =& \, A'' + \frac{\ddot A}{f} + A'^2 + \frac{\dot A^2}{f} + (D-3) \left( A'C' + \frac{\dot A\dot C}{f} + \frac{A'}{r} + \frac{f'C'}{2f} \right) + \frac{3f'A'}{2f} \, ,	
			\tag{\theequation a} \label{eq:field_eqnsA} \\
	0 =& \, B'' + \frac{\ddot B}{f} - (D-3) \left( A'C' + \frac{\dot A\dot C}{f} + \frac{A'}{r} + \frac{f'C'}{2f} \right) 	 + \frac{f'B'}{2f}								    \nonumber \\
	   & \, - \frac{(D-3)(D-4)}{2r^2} \left( \frac{1 - \E^{2B-2C}}{f}  + r^2C'^2 + 2rC' + \frac{r^2\dot C^2}{f} \right) \, ,		              
	   		\tag{\theequation b} \label{eq:field_eqnsB}	\\
	0 =& \, C'' + \frac{\ddot C}{f} + A'C' + \frac{\dot A \dot C}{f} + \frac{A'}{r} + \frac{f'C'}{f}  																		    \nonumber \\
	   & \, + \frac{(D-4)}{r^2} \frac{ (1-\E ^{2B-2C})}{f}+ (D-3) \left( C'^2 + \frac{2C'}{r} + \frac{\dot C^2}{f} \right) \, .  			  
	   		\tag{\theequation c} \label{eq:field_eqnsC}	
\end{align}
Note that Einstein's equations actually give five independent equations. The missing two ones, $G^r_z=0$ and $G^r_r-G^z_z=0$, can be regarded as constraints which are satisfied globally by solutions of~\eqref{eq:field_eqns} if appropriate boundary conditions are imposed (this is Wiseman's ``constraint rule''~\cite{Wiseman:2002zc}, see~\cite{Kleihaus:2006ee} for the choice of coordinates used in \eqref{eq:line_element}). 

The domain of integration is $\mathcal{G} = \{ (r,z) : r_0 \leq r < \infty \, , ~ 0 \leq z \leq L/2 \}$. Here we have incorporated reflection symmetry with respect to the coordinate line $z=L/2$, which is inherent to our problem. Hence it suffices to restrict the $z$-coordinate accordingly. 

The following boundary conditions arise:
\begin{align}
	0 =&      A =      B =      C \quad \text{at} \quad r\to\infty \, , \label{eq:BC_infinity} \\
	0 =& \dot A = \dot B = \dot C \quad \text{at} \quad z = 0 \quad \text{and} \quad z = L/2 \, . \label{eq:BC_poles} 
\end{align}
The conditions~\eqref{eq:BC_infinity} imply asymptotic ``flatness'', while the equations~\eqref{eq:BC_poles} result from the periodicity and reflection symmetry in $z$.\footnote{Here, ``flat'' means Minkowski space times a circle.} On the horizon $r=r_0$ we can impose a ``constant temperature condition'' (cf. section~\ref{subsec:Horizon_quantities}) by
\begin{equation}
	0 = \dot A - \dot B \quad \text{at} \quad r=r_0 \, . 
	\label{eq:BC_const_T}
\end{equation}
Apparently, condition~\eqref{eq:BC_const_T} contains an undetermined constant of integration. We introduce the freely specifiable value  $\beta _\text{c}$ by requiring that
\beq
	0 = \E ^{-2B} - \beta _\text{c}  \quad \text{at} \quad (r,z)=(r_0,L/2) \, ,
	\label{eq:BC_crit_point}
\eeq
through which $B$ is prescribed at one point on the horizon and the constant in question is fixed. It turns out that $\beta _\text{c}$ (having no significant physical meaning) is a useful parameter which controls the transition of our solutions along the NBS branch towards the critical regime. Beginning with $\beta _\text{c}=1$ for a UBS, it monotonically decreases and reaches zero for the limiting pinch-off solution.  

There are two additional conditions which result from the field equations~\eqref{eq:field_eqns} in the degenerate limit  $r\to r_0$.
Instead of using these conditions, we follow~\cite{Kleihaus:2006ee} and introduce a modified radial coordinate $\varrho$ via $r/r_0=\sqrt{\varrho^2+1}$ and obtain, as regularity requirements, that the derivatives with respect to $\varrho$ vanish on the horizon. Our additional two boundary conditions are therefore
\beq
	0 = A_{,\varrho} = C_{,\varrho} \quad \text{at} \quad r = r_0 \quad \text{(or } \varrho = 0) \, .
	\label{eq:BC_horizon_xi}
\eeq
The conditions~\eqref{eq:BC_infinity} to~\eqref{eq:BC_horizon_xi} are in accordance with the aforementioned constraint rule~\cite{Wiseman:2002zc}. Note that also $B_{,\varrho}=0$ holds on the horizon, but we require instead equation~\eqref{eq:BC_const_T}. As a consequence of the validity of the constraints it then follows that $B_{,\varrho}$ vanishes for $r=r_0$.

Finally we remark that a unique solution to the equations \eqref{eq:field_eqns}--\eqref{eq:BC_horizon_xi} is obtained by scaling physical quantities in terms of appropriate powers of $r_0$ and fixing $L/r_0$ (we take $L/r_0 = L_{\text{GL}}/r_0$, cf.~\eqref{eq:L_GL}), and, moreover, by prescribing a value for $\beta _{\text{c}}\in [0,1]$. 

Note that although not all of the constraint conditions enter our numerical scheme (which solves the system \eqref{eq:field_eqns}--\eqref{eq:BC_horizon_xi}), they can be used {\em a posteriori} to analyze consistency and accuracy of the solution.

\subsection{Asymptotics}
\label{subsec:Asymptotics}

We proceed with a more elaborate discussion of the asymptotics of the metric functions. The equation \eqref{eq:BC_infinity} can be refined as follows~\cite{Kol:2003if}:
\refstepcounter{equation}\label{eq:asymptotics}
\begin{align}
	\lim_{r\to\infty}r^{D-4}A&= A_{\infty}r_0^{D-4}  \, , 
			\tag{\theequation a} \label{eq:asymptoticsA} \\
	\lim_{r\to\infty}r^{D-4}B&= B_{\infty}r_0^{D-4}  \, , 
			\tag{\theequation b} \label{eq:asymptoticsB} \\
	\lim_{r\to\infty}g_D(r)C&=C_{\infty}r_0\quad\mbox{with} \quad
	g_D(r)= \begin{cases}
  			r(\log r)^{-1} & \text{if } D=5       \, , \\
  			r                & \text{if } D\geq 6 \, , 
		  \end{cases}	
		  	\tag{\theequation c} \label{eq:asymptoticsC} 
\end{align}
and constants $A_\infty, B_\infty$ and $C_\infty$. Note the logarithmic term in the asymptotics of the function $C$ in five spacetime dimensions which calls for a careful treatment if one wishes to produce highly accurate numerical results.\footnote{One might think of the simple coordinate transformation $s=1/r$ which compactifies the infinite region. In terms of $s$, the fall-off in $C$ reads: $C\sim\log r/r = -s\log s$. Obviously, no derivatives (with respect to $s$) of this term exist at infinity ($s=0$). This leads to a poor convergence of any spectral approximation of this function. We provide an alternative approach in subsection~\ref{subsubsec:Treatment_of_the_asymptotics_5D}.}

From the asymptotic values we can extract two charges~\cite{Kol:2003if,Harmark:2003dg}, namely the black string mass
\beq
	M = M_0 \left( 1 - 2A_\infty - \frac{2}{D-3}B_\infty \right) \quad \text{with} \quad M_0 = \frac{(D-3) r_0^{D-4} L \Omega _{D-3} }{16\pi G}
	\label{eq:mass}
\eeq
and the relative tension
\beq
	n = n_0 \, \frac{1 - 2A_\infty - 2B_\infty  (D-3)}{1 - 2A_\infty - 2 B_\infty / (D-3)}  \quad \text{with } \quad n_0 = \frac{1}{D-3} \, .
	\label{eq:tension}
\eeq
Here, $\Omega _{D-3}$ is the surface area of a unit $(D-3)$-sphere and $G$ is Newton's constant of gravitation. The quantities $M_0$ and $n_0$ describe the corresponding values obtained for the UBS.
\subsection{Horizon quantities}
\label{subsec:Horizon_quantities}
In this section we define relevant geometric as well as thermodynamic quantities which characterize the black string horizon. From the line element (\ref{eq:line_element}) we read off a $z$-dependent circumferential horizon radius 
\beq
	\label{eq:areal_radius}
	R(z) = r_0\,\E ^{C(r_0,z)} \, ,
\eeq
which is referred to as the ``horizon areal radius''~\cite{Sorkin:2006wp}. The parameters $R_{\rm max}$ and  $R_{\rm min}$ introduced in eq.~\eqref{eq:lambda} are given by the corresponding extremal values, i.e.:
\beq
	\label{eq:Rmin_Rmax}
	R_{\rm min}=\min_{0\le z\le L/2} R(z),\qquad R_{\rm max}=\max_{0\le z\le L/2} R(z).
\eeq
Next, the proper length of the compact dimension on the horizon is defined by
\beq
	L_\mathcal{H} = \int _0^L \E ^{B(r_0,z)} \,\D z \,.
	\label{eq:L_H}
\eeq
Apart from this, we introduce in the usual manner the following thermodynamic quantities: (i) the black string temperature $T$, which is proportional to the constant surface gravity on the horizon,\footnote{Note that a $z$-independent temperature means that the difference between $A$ and $B$ is constant on the horizon, which is satisfied by our boundary condition~\eqref{eq:BC_const_T}.}
\beq
	T = T_0 \, \E ^{A(r_0,z)-B(r_0,z)} \quad \text{with} \quad T_0 = \frac{D-4}{4\pi r_0},
	\label{eq:T}
\eeq
and (ii) its entropy, which is proportional to the horizon area,
\beq
	S = \frac{S_0}{L} \int _0^L \E ^{B(r_0,z) + (D-3) C(r_0,z)} \,\D z \quad \text{with} \quad S_0 = \frac{r_0^{D-3} L \Omega _{D-3}}{4G} \, .
	\label{eq:S}
\eeq
Similar to the above, $T_0$ and $S_0$ denote corresponding values of the UBS.

Together with the mass and the relative tension, temperature and entropy obey Smarr's relation~\cite{Kol:2003if,Harmark:2003dg}
\beq
	TS = \frac{D-3-n}{D-2}M \, ,
	\label{eq:Smarr}
\eeq
as well as the first law of black hole thermodynamics
\beq
	\D M = T \D S + \frac{nM}{L} \D L \, .
	\label{eq:1stlaw}
\eeq
Since these formulas relate asymptotic charges with horizon values, they can be used as non-trivial tests of consistency and accuracy of the numerical scheme, cf. section~\ref{subsec:Accuracy}. In the following we fix $L/r_0$ by $L_{\text{GL}}/r_0$ (cf.~\eqref{eq:L_GL}), which means that the first law reduces to the form $\D M = T \D S$.

\section{Numerical implementation}
\label{sec:Numerical_implementation}
The basis of our numerical scheme is the expansion of any function $g:[a,b]\to \mathbb{R}$ in terms of Chebyshev polynomials $T_k(y) = \cos [k \arccos (y)]$, where $y\in [-1,1]$ and
\beq
	g(x) \approx \sum _{k=0}^{N-1} c_k T_k\left( \frac{2x-b-a}{b-a} \right) \, .
	\label{eq:spectral_expansion}
\eeq 
Considering the functions' values on Lobatto grid points (which include the boundaries $a$ and $b$)
\beq
	x_k = (b-a) \sin ^2\left[ \frac{\pi k}{2(N-1)} \right] + a \quad \text{with} \quad k=0,1,\ldots ,N-1  \, ,
	\label{eq:Lobatto}
\eeq
we can calculate the Chebyshev coefficients $c_k$ as well as approximate derivatives of $g$ by standard pseudo-spectral schemes. For a given expansion order $N$ the accuracy of the approximation~\eqref{eq:spectral_expansion} depends crucially on the fall-off of the coefficients $c_k$, which is governed (roughly speaking) by smoothness properties of the underlying function $g$. Thus, our strategy consists of a reformulation of the problem in order to obtain rapidly decaying coefficients of the solution when considered on appropriate domains. We provide a detailed discussion of this treatment in the upcoming sections \ref{subsec:6D_implementation} and \ref{subsec:5D_implementation}.

Before moving on we note that we apply the Newton-Raphson method for solving the discretized system describing the collection of non-linear partial differential equations and boundary conditions ~\eqref{eq:field_eqns}--\eqref{eq:BC_horizon_xi}. In the several iterative steps of this scheme a linear system involving a Jacobian matrix has to be solved. This is done by means of the so-called BiCGSTAB method~\cite{barrett1994templates}, which we endow with a preconditioner that utilizes a finite difference representation of the Jacobian. We finally remark that the linear system arising within the preconditional step is solved efficiently with the help of a band matrix decomposition algorithm~\cite{Press:2007:NRE:1403886}.

\subsection{Non-uniform black strings in 6 spacetime dimensions}
\label{subsec:6D_implementation}

We start with the description of the numerical scheme for the construction of solutions describing NBS in six spacetime dimensions. As mentioned above, these solutions are not plagued with logarithmic terms. The introduction of a modified radial coordinate $\chi$ via
\beq
	\frac{r_0}{r} = 1-(1-\chi)^2 = \chi (2-\chi ) \, ,
	\label{eq:chi}
\eeq
realizes on the one hand a compactification of the infinite domain and yields on the other hand the regularity conditions \eqref{eq:BC_horizon_xi}, 
which are realized by requiring that the derivatives with respect to $\chi$ vanish on the horizon. Now, a straightforward implementation of the equations~\eqref{eq:field_eqns} to~\eqref{eq:BC_horizon_xi} in terms of a pseudo-spectral method on a single compactified computational domain according to the scheme described above does not provide  us with a satisfactory accuracy, not even for solutions near the UBS with $\lambda\ll 1$. In the following we describe a more sophisticated approach which consists of a special treatment of the solution's behavior for $r\to\infty$ (relevant for all $\lambda\in(0,\infty)$, i.e.~also for small horizon deformation) as well as a technique to handle the critical regime where the horizon deformation becomes very large ( $\lambda\gg 1$).

\subsubsection{Treatment of the asymptotics}
\label{subsubsec:Treatment_of_the_asymptotics_6D}

A detailed analysis (see appendix~\ref{appendix:subsec:6D_perturbations}) of linear perturbations around the UBS leads us to the following ansatz for our metric potentials:
\refstepcounter{equation} \label{eq:splitAnsatz6D}
\begin{align}
	A =  &  A_0(r) \left( \dfrac{r_0}{r} \right) ^2            + \, A_1(r ,z) \, \cos\left( \tfrac{2\pi}{L}z\right) \, \E ^{-2\pi r/L} \, \left( \dfrac{r_0}{r} \right) ^{3/2} \, , 	
		\tag{\theequation a} \label{eq:splitAnsatz6DA}  \\
	B =  &  B_0(r) \left( \dfrac{r_0}{r} \right) ^2            + \, B_1(r ,z) \, \cos\left( \tfrac{2\pi}{L}z\right) \, \E ^{-2\pi r/L} \, ,					                           
		\tag{\theequation b} \label{eq:splitAnsatz6DB} \\ 
	C =  &  C_0(r) \left( \dfrac{r_0}{r} \right) \hphantom{^2} + \, C_1(r ,z) \, \cos\left( \tfrac{2\pi}{L}z\right) \, \E ^{-2\pi r/L} \, \left( \dfrac{r_0}{r} \right) \, . 
		\tag{\theequation c} \label{eq:splitAnsatz6DC}  					  
\end{align}
This ansatz describes a decomposition into a part which depends only on the radial coordinate and is dominant near infinity, and a remainder with specified leading terms in the radial direction. 
According to \eqref{eq:asymptotics}, we can directly read off the asymptotic charges $M$ and $n$ from the boundary values $A_0|_{r\to\infty} =A_\infty$ and  $B_0|_{r\to\infty} =B_\infty$. Also we have $C_0|_{r\to\infty} =C_\infty$. Note that there is a one-to-one map between $\{A_0, A_1\}$ and $A$ as they follow from
\begin{align}
	A_0(r)   = & \left(\dfrac{r_0}{r} \right)^{-2} A(r,z)|_{z=L/4} \, ,  
 		\label{eq:A0_of_A} \\
	A_1(r,z) = & \frac{A(r,z)-A(r,z)|_{z=L/4}}{\cos\left( \tfrac{2\pi}{L}z\right)} \, \E ^{2\pi r/L} \, \left( \dfrac{r_0}{r} \right) ^{-3/2} \, ,
		\label{eq:A1_of_A}
\end{align}
and similarly for $B$ and $C$. 

Now  observe the half-integer power of $r$ in \eqref{eq:splitAnsatz6D} which, expressed in terms of $\chi$ (see \eqref{eq:chi}) is non-smooth, $r^{-3/2}\sim  \chi^{3/2}$. This consideration leads us, for the six-dimensional case, to the introduction of yet another radial coordinate $\xi\in [0,1]$ via:
\beq
	\frac{r_0}{r} = \left[ 1 - (1-\xi )^2 \right] ^2 = \xi ^2 (2-\xi )^2 \, .
	\label{eq:xi6D}
\eeq
Through the introduction of $\xi$ we achieve (i) compactification of the infinite domain, (ii) vanishing $\xi$-derivatives on the horizon and (iii) regularization of half-integer powers of $r_0/r$ near infinity. The coordinate values  $\xi =0$ and $\xi=1$ correspond to infinity and to the horizon respectively. Note that the term $\E ^{-2\pi r/L}$ is not analytic with respect to $\xi$ at $\xi = 0$, because all of its $\xi$ derivatives vanish at this point (the same holds if it is expressed in terms of $\chi$).

The fall-off of the spectral coefficients with respect to the angular direction $z$ can be enhanced by introducing the new coordinate $u\in [-1,1]$:
\beq
	u = \cos\left( \tfrac{2\pi}{L}z\right) \, ,
	\label{eq:u}
\eeq
where $z=0$ corresponds to $u=1$, $z=L/4$ to $u=0$ and $z=L/2$ to $u=-1$.\footnote{\label{note:oddN}We note that the ansatz~\eqref{eq:splitAnsatz6D} privileges the use of an odd number of grid points with respect to the $u$-direction, since then the central point $u=0$ (corresponding to $z=L/4$) is contained within the Lobatto grid~\eqref{eq:Lobatto}.} More specifically, in higher orders of perturbation theory, only powers of $u$ appear, and hence we expect a rapid convergence of the spectral coefficients in this coordinate at least in the vicinity of the UBS solution. We note that a Chebyshev expansion in $u$ is basically equivalent to an even Fourier expansion in $z$, cf. eq.~\eqref{eq:spectral_expansion}. 

In the following we present asymptotic boundary conditions for the auxiliary functions which constitute the metric components through~\eqref{eq:splitAnsatz6D}. At first, we derive conditions for the $z$-independent functions $A_0(\xi )$, $B_0(\xi )$ and $C_0(\xi )$ through the analysis of a power series in terms of $\xi$ of the field equations~\eqref{eq:field_eqns} at $u=0$ (i.e. $z=L/4$) about $\xi =0$. To zeroth order, we obtain $A_{0,\xi}=0$ from \eqref{eq:field_eqnsA} and the condition $C_{0,\xi}=0$ from both \eqref{eq:field_eqnsB} and \eqref{eq:field_eqnsC}. Taking these findings into account for the first expansion order, we get
\beq\label{eq:asympTaylor1st_1}0 = 3 A_{0,\xi\xi} - 24 (1-2 A_\infty )C_\infty \eeq
from \eqref{eq:field_eqnsA} and 
\beq\label{eq:asympTaylor1st_2} C_{0,\xi\xi} = -4(2 A_\infty +4 B_\infty + C_\infty ^2)\eeq
again from both \eqref{eq:field_eqnsB} and \eqref{eq:field_eqnsC}. At the second order we observe that by virtue of the preceding relations also $B_{0,\xi}=0$ holds. Now, the numerical scheme based on the boundary conditions of vanishing $\xi$-derivatives for the three functions $A_0$, $B_0$ and $C_0$ at $\xi=0$ does work well but yields unsatisfactory accuracy. This is due to the fact that the condition $B_{0,\xi}=0$ arises in the course of the power law expansion only at second order which affects the rounding error of internal computations. Therefore, in order to restore high accuracy we utilize the condition $C_{0,\xi}=0$ and perform another decomposition
\beq
	C_0(\xi) = C_{\infty} + \xi ^2 \, C_{01}(\xi) \, .
	\label{eq:C0split} 
\eeq
Now, reconsidering our power law expansion at second order, we find besides $B_{0,\xi}=0$ two further conditions, of which one reads:
\beq\label{eq:asympTaylor2nd}C_{0,\xi\xi\xi} =12 ( 2 A_\infty + 4 B_\infty + C_\infty ^2 ).\eeq
Incorporating \eqref{eq:C0split} we finally arrive at the following collection of boundary conditions for the three functions $A_0$, $B_0$ and $C_0$ at $\xi=0$:
\beq
\begin{array}{lllllll}
	A_{0,\xi}                                      &=& 0 &,  & B_{0,\xi} &=& 0 \, ,  \\[2mm]
	C_{01} + 2(2A_\infty + 4B_\infty+C_{\infty}^2) &=& 0 &,  & C_{01} + C_{01,\xi} &=& 0\, .
\end{array}	\label{eq:asympBC0}
\eeq 
Here, third and fourth condition are equivalent to \eqref{eq:asympTaylor1st_2} and \eqref{eq:asympTaylor2nd}. Note that besides the function $C_{01}$ the value $C_\infty$ appears as additional unknown in our numerical scheme which means that we have to take {\em four} (instead of three) conditions into account. Eq.~\eqref{eq:asympTaylor1st_1} is neglected in the set of boundary conditions but emerges {\em a posteriori} as property of the final solution.

For the $z$-dependent functions $A_1$, $B_1$ and $C_1$ the field equations at infinity, $\xi =0$, give the conditions
\beq
	u(1-u^2)X_{1,uu} - (2-3u^2) X_{1,u} - 2 X_{1,u}|_{u=0} = 0 \, ,
	\label{eq:asympBCode}
\eeq
where $X_1(u) = \{ A_1(\xi ,u), B_1(\xi ,u) , C_1(\xi ,u) \} |_{\xi =0}$. The only regular solution of this ordinary differential equations is $X_{1,u}(u)=0$ and therefore we obtain the following boundary conditions at $\xi =0$
\beq
	A_{1,u} = B_{1,u} = C_{1,u} = 0 \, ,
	\label{eq:asympBC1}
\eeq
or in other words, these functions take  $z$-independent constant values at infinity which are unknown {\em a priori}. This comes not as a surprise, since the $z$-dependent modes of the functions $A_1$, $B_1$ and $C_1$ carry again the factor $\E ^{-2\pi r/L}$ and are therefore rapidly decreasing, which can be seen explicitly through an expansion of the perturbation theory to higher orders.\footnote{\label{note:rapdec}By default the term ``rapidly decreasing'' refers to functions which tend asymptotically to zero faster than any inverse power of $r$.} Accordingly, we can derive additional asymptotic conditions from a power law expansion in terms of $\xi$ of the corresponding equations at $\xi =0$, thereby neglecting the $z$-dependent modes of $A_1$, $B_1$ and $C_1$. The leading order yields
\beq
	A_{1,\xi} = 0 \, , \quad B_{1,\xi} = 0 \, , \quad B_1 + 2\pi \frac{r_0}{L}\, C_1 = 0 \, ,
	\label{eq:asympBC10}
\eeq
which together with~\eqref{eq:asympBC1} constitute the set of asymptotic conditions for the functions $A_1$, $B_1$ and $C_1$. In our numerical scheme, we require~\eqref{eq:asympBC1} at all grid points $(\xi,u)=(0,u)$ with $u\ne 0$, while at $(\xi,u)=(0,0)$ the conditions~\eqref{eq:asympBC10} are enforced.

\subsubsection{Decomposition of the numerical domain}
\label{subsubsec:Decomposition_of_the_numerical_domain_6D}

\begin{figure}[!ht]
	\includegraphics[scale=1]{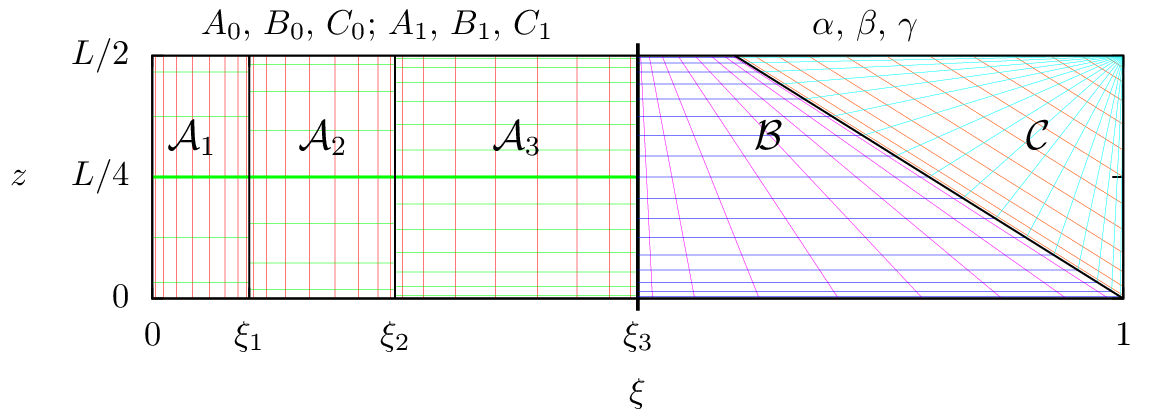}
	\caption{Division of the integration domain into several subdomains for $D=6$. The asymptotic region $\xi\leq\xi_I$ is split up into $I$ rectangles (here $I=3$) with different resolutions regarding the 	$z$-direction. In this region the numerical algorithm solves for the auxiliary functions 	$A_0$, $B_0$, $C_0$, $A_1$, $B_1$ and $C_1$ which constitute through~\eqref{eq:splitAnsatz6D} the metric 			functions $A,B$ and $C$. Note that the one-dimensional functions $A_0$, $B_0$ and $C_0$ are 			considered on the $z=L/4$-line	(drawn more boldly, see also footnote~\ref{note:oddN}). Also, for 		$\xi\leq\xi_I$ we express the functions $A_1$, $B_1$ and $C_1$ in terms of $\xi$ and $u$ (cf.~\eqref{eq:xi6D} and~\eqref{eq:u}).
	The near-horizon region $\xi\ge\xi_I$ is split up into the trapezoidal subdomain $\mathcal B$ and the triangular subdomain $\mathcal C$ which is 
	divided up further, in order to resolve steep gradients at the critical point $(\xi, z ) = (1,L/2)$, 
	see fig.~\ref{fig:int_domain_triangle_0}. In $\mathcal B$ and $\mathcal C$, the numerical algorithm solves for the auxiliary functions 
	$\alpha $, $\beta $ and $\gamma$ which are related through~\eqref{eq:albega} to $A,B$ and $C$.}
	\label{fig:int_domain}
\end{figure}

In the course of the development of the code it has turned out that an appropriate splitting of the domain of integration is essential to render highly accurate solutions. With respect to the coordinates $(\xi ,u)$ we denote the entire domain by $\mathcal G = \{ (\xi ,u): 0\leq \xi \leq 1 \, , ~ -1 \leq u \leq 1\}=[0,1]\times[-1,1]$. In a first step we divide up the asymptotic region, more concretely, we choose some $\xi_I$ with $0<\xi_I<1$ and decompose the subset $[0,\xi_I]\times[-1,1]$ into $I$ subdomains $\mathcal A_i = [\xi_{i-1},\xi_i]\times[-1,1]$, where $i = 1,2,\ldots ,I$ and $0=\xi_0 < \xi _1 < \ldots < \xi _I$. The benefit is twofold. On the one hand we obtain a rapid decay of the spectral coefficients with respect to  the $\xi$-direction in each $\mathcal A_i$ if the boundary values $\xi_i$ are chosen appropriately. In comparison to a spectral expansion on $\mathcal G$ we achieve a substantial improvement of the accuracy since we can take into account the non-analytic behavior in \eqref{eq:splitAnsatz6D} by choosing narrow domains in the vicinity of $\xi=0$, see fig.~\ref{fig:int_domain}. On the other hand, we are able to use different resolutions with respect to the $u$-direction in each $\mathcal A_i$. So near $\xi =0$, where the $u$-dependence is rapidly suppressed, we may take a low resolution, while close to the horizon $\xi = 1$ much greater resolutions are needed. From a computational point of view, this saves a lot of memory space as well as computing time and increases, moreover, the accuracy of the solutions. Note that the implementation of non-uniform resolutions requires interpolation at the boundaries between two subdomains to impose equality of the functions' values and their normal derivatives.

For the remaining subset $[\xi_I,1]\times[-1,1]$ we take into account that in the limit $\lambda\to\infty$ the functions $A$, $B$ and $C$ diverge at the critical point $(\xi ,u) = (1,-1)$ (corresponding to $(r, z ) = (r_0,L/2)$).\footnote{Note that this is exactly the spatial grid point where the horizon is supposed to pinch-off in this limit.} In order to tackle this situation we introduce the auxiliary functions, $\alpha$, $\beta$ and $\gamma$ through
\beq
	\alpha = \E ^{-2A} \, , \quad \beta = \E ^{-2B} \, , \quad \gamma = \E ^{2C} \, .
	\label{eq:albega}
\eeq
In the limit $\lambda\to\infty$ the values of $\alpha,\beta$ and $\gamma$ tend to zero at the critical point. This point is, however, plagued with steep gradients when considering large $\lambda$, in particular with respect to the $u$-direction. We recognize that in the critical regime the benefits of the coordinate $u$ are lost, and we return therefore near the horizon back to the coordinate $z$ (elsewhere we can still profit from the properties of $u$).\footnote{In our numerical implementation we use dimensionless quantities, i.e.~the calculations are carried out with respect to $z/L$.}

\begin{figure}[!ht]
	\includegraphics[scale=0.96]{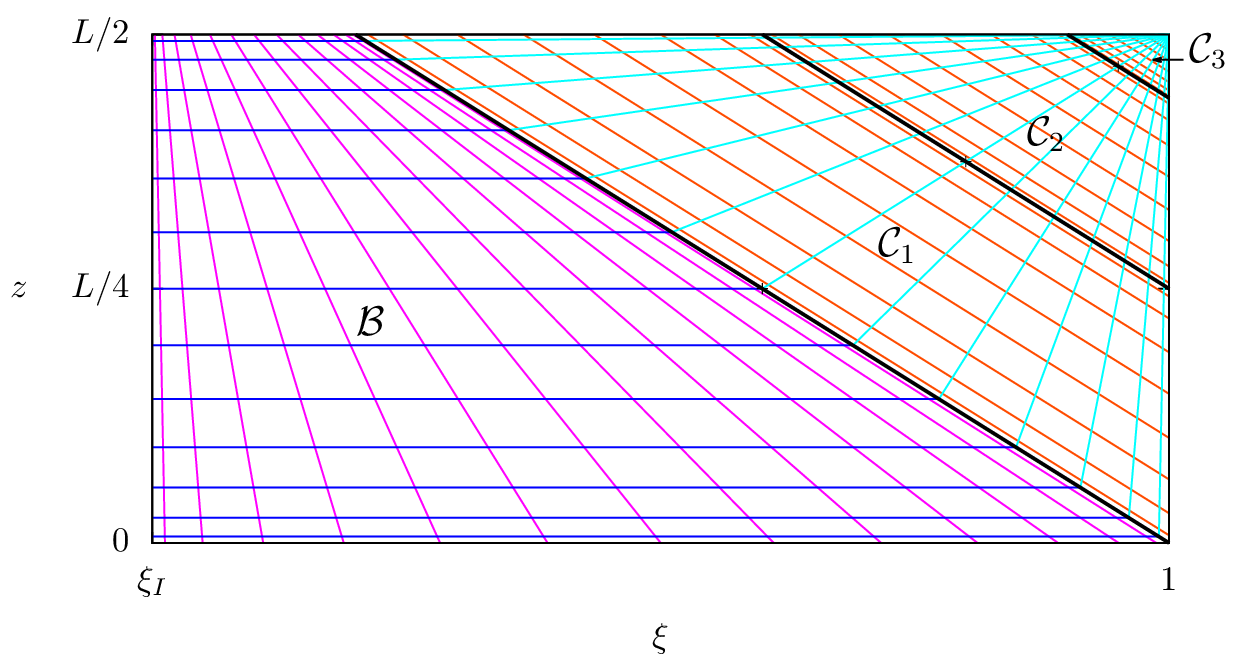}
	\caption{In the case $D=6$, the region $\xi \geq \xi _I$ is split up into the trapezoidal subdomain $\mathcal B$ and the triangular subdomain $\mathcal C$ which is divided up further into $J$ subdomains (here $J=3$). See fig.~\ref{fig:int_domain} and appendix~\ref{appendix:sec:Resolving_the_critical_point} for details.}
	\label{fig:int_domain_triangle_0}
\end{figure}

Still, in the coordinates $(\xi,z)$ we observe for large $\lambda$ clearly pronounced peaks of the functions $\alpha$, $\beta$ and $\gamma$ at the critical point. We resolve these peaks by splitting the domain $[\xi_I,1]\times [0,L/2]$ into a trapezoidal subdomain $\mathcal B$ and a triangular subdomain $\mathcal C$, see fig.~\ref{fig:int_domain}. The latter one is then further subdivided into $J$ subdomains $\mathcal C_j$ ($j=1,\ldots, J$) placed around the critical point, see fig.~\ref{fig:int_domain_triangle_0}. The corresponding coordinate transformations which provide the mappings of these subdomains onto rectangles are given in appendix \ref{appendix:sec:Resolving_the_critical_point}.


\subsection{Non-uniform black strings in 5 spacetime dimensions}
\label{subsec:5D_implementation}
\subsubsection{Treatment of the asymptotics}
\label{subsubsec:Treatment_of_the_asymptotics_5D}

Again a sophisticated analysis of linear perturbations around the UBS, to be conducted in appendix~\ref{appendix:subsec:5D_perturbations}, suggests the following ansatz near infinity in 5 spacetime dimensions:
\refstepcounter{equation} \label{eq:splitAnsatz5D}
\begin{align}	
	A &=  \hphantom{-} A_0(r) \, \dfrac{r_0}{r} \hphantom{\log \dfrac{r_0}{r}} + \, A_1(r ,z) \, \cos\left( \tfrac{2\pi}{L}z\right) \, \left( \dfrac{r_0}{r} \right) ^4 \, ,  
		\tag{\theequation a} \label{eq:splitAnsatz5DA} \\
	B &=  \hphantom{-} B_0(r) \, \dfrac{r_0}{r} \hphantom{\log \dfrac{r_0}{r}} + \, B_1(r ,z) \, \cos\left( \tfrac{2\pi}{L}z\right) \, \left( \dfrac{r_0}{r} \right) ^4 \, ,							         
		\tag{\theequation b} \label{eq:splitAnsatz5DB} \\ 
	C &=            -  C_0(r) \, \dfrac{r_0}{r}            \log \dfrac{r_0}{r} + \, C_1(r ,z) \, \cos\left( \tfrac{2\pi}{L}z\right) \, \left( \dfrac{r_0}{r} \right) ^4  \, . 					         
		\tag{\theequation c} \label{eq:splitAnsatz5DC} 			  
\end{align}
Like in 6 spacetime dimensions (cf.~\eqref{eq:splitAnsatz6D}) we split the functions into an asymptotically dominant part which only depends on $r$ and a secondary term depending on both coordinates, $r$ and $z$. The secondary terms are rapidly decreasing (cf. footnote~\ref{note:rapdec}) asymptotically. However, in $D=5$ spacetime dimension, we refrain from extracting the corresponding exponential factor from the secondary terms, see appendix~\ref{appendix:subsec:5D_perturbations}. Comparing~\eqref{eq:splitAnsatz5D} with~\eqref{eq:asymptotics} we again find $A_0|_{r\to\infty} =A_\infty$, $B_0|_{r\to\infty} =B_\infty$ and $C_0|_{r\to\infty} =C_\infty$.

Similar to the case $D=6$, we consider the ansatz~\eqref{eq:splitAnsatz5D} only in regions far away from the horizon. This means that the evidently incorrect condition $C(r_0,L/4)=0$ (cf.~\eqref{eq:splitAnsatz5DC}) does not appear in our numerical set up. Also, in these regions we again benefit from the coordinate $u$ defined in~\eqref{eq:u}. 

In the radial direction we work with the coordinate $\chi$ defined in~\eqref{eq:chi}. Observe that in \eqref{eq:splitAnsatz5D} no half-integer powers of $r_0/r$ appear which eliminates the need to introduce the coordinate $\xi$ via \eqref{eq:xi6D}. Such terms as well as logarithmic expressions (like $\log ^m (r_0/r)$) are suppressed by the rapid decay of the functions $A_1$, $B_1$ and $C_1$. Consequently, we expect a sub-geometric (stronger than algebraic) convergence of the spectral coefficients of $A_1$, $B_1$ and $C_1$ with respect to $\chi$.\footnote{We follow the terminology of Boyd~\cite{boyd2001chebyshev}.} In contrast, the asymptotics of the functions $A_0$, $B_0$ and $C_0$ contains logarithmic terms in $\chi$. In order to guarantee globally high accuracy, we consider these functions in terms of another radial coordinate $\eta\in [0,1]$ given through 
\beq
	\eta\in[0,1],\,\chi\in[0,\chi_I]:\qquad
	\eta=\frac{1}{1-\log(\chi/\chi_I)}\quad\Leftrightarrow\quad\chi = \chi _I \, \E ^{1-\frac{1}{\eta}} \, ,
	\label{eq:eta}
\eeq
where $0\leq\chi _I\leq 1$. Apparently, the asymptotic boundary $\chi=0$ is obtained for $\eta\to 0$, whereas $\eta=1$ corresponds to $\chi=\chi_I$ which, similar as in the case $D=6$ separates the asymptotic from the near-horizon domain, see fig.~\ref{fig:int_domain_5D}. Note that through \eqref{eq:eta} expressions of the kind $\chi ^l\log^m \chi$ ($l,m>0$) are infinitely many times differentiable (but not analytic) with respect to $\eta$ at $\eta =0$. In this way we assure sub-geometric convergence of the spectral coefficients of $A_0$, $B_0$ and $C_0$. 

We summarize that the functions $A_0$, $B_0$ and $C_0$ are considered at gridpoints with specifically prescribed values of the coordinate $\eta$ (cf.~\eqref{eq:eta}), while the functions $A_1$, $B_1$ and $C_1$ are evaluated at particularly chosen coordinate values on the $(\chi ,u)$-grid (see \eqref{eq:chi}, \eqref{eq:u}). In the numerical set up, we therefore need to apply interpolation techniques, in order to obtain a mapping between the different coordinate grids. As this is done only at the coordinate line with fixed value $u=0$ (i.e.~$z=L/4$), the corresponding computational costs barely come into account. The two different radial grids are displayed in fig.~\ref{fig:int_domain_5D}.

We now provide the corresponding asymptotic boundary conditions. From the relevant equations for the functions $A_0$, $B_0$ and $C_0$, derived from~\eqref{eq:field_eqns} in terms of the coordinate $\eta$ and considered at $u=0$ (i.e. $z=L/4$), we obtain in the limit $\eta\to 0$:
\beq
	A_{0,\eta} = 0 \, , \quad B_{0,\eta} = 0 \, , \quad A_{\infty} + 2B_{\infty} - C_{\infty} = 0 \, .
	\label{eq:asympBC5Dradial}
\eeq
This set of conditions turns out to work well and yields satisfactory precision. Hence, an additional decomposition of the function $C_0$ (as was done for $D=6$) is not necessary. Finally, the rapid decay of the functions $A_1$, $B_1$ and $C_1$ simply implies the conditions 
\beq
	A_1 = B_1 = C_1 = 0,
	\label{eq:asympBC5Dz}
\eeq
to be satisfied asymptotically at $\chi=0$.

\subsubsection{Decomposition of the numerical domain}
\label{subsubsec:Decomposition_of_the_numerical_domain_5D}

Similarly to the case $D=6$ we split up the integration domain  $\mathcal G = \{ (\chi ,u): 0\leq \chi \leq 1 \, , ~ -1 \leq u \leq 1\}=[0,1]\times[-1,1]$ into several subdomains (see~\eqref{eq:chi} and~\eqref{eq:u}). The region $[0,\chi_I]\times[-1,1]$, in which we utilize the ansatz~\eqref{eq:splitAnsatz5D}, is divided up into $I$ subdomains $\mathcal A_i = [\chi_{i-1},\chi_i]\times[-1,1]$, where $i = 1,2,\ldots ,I$ and $0=\chi_0 < \chi _1 < \ldots < \chi _I$ for the functions $A_1$, $B_1$ and $C_1$. Likewise, the functions $A_0$, $B_0$ and $C_0$ are considered with respect to the radial coordinate $\eta$ (see~\eqref{eq:eta}) on several intervals $[\eta _{i-1},\eta _i]$ where $0=\eta _0<\eta _1< \ldots <\eta _I=1$. The benefits of this approach are the same as in the case $D=6$, discussed in section~\ref{subsubsec:Decomposition_of_the_numerical_domain_6D}. 

\begin{figure}[!ht]
	\includegraphics[scale=1]{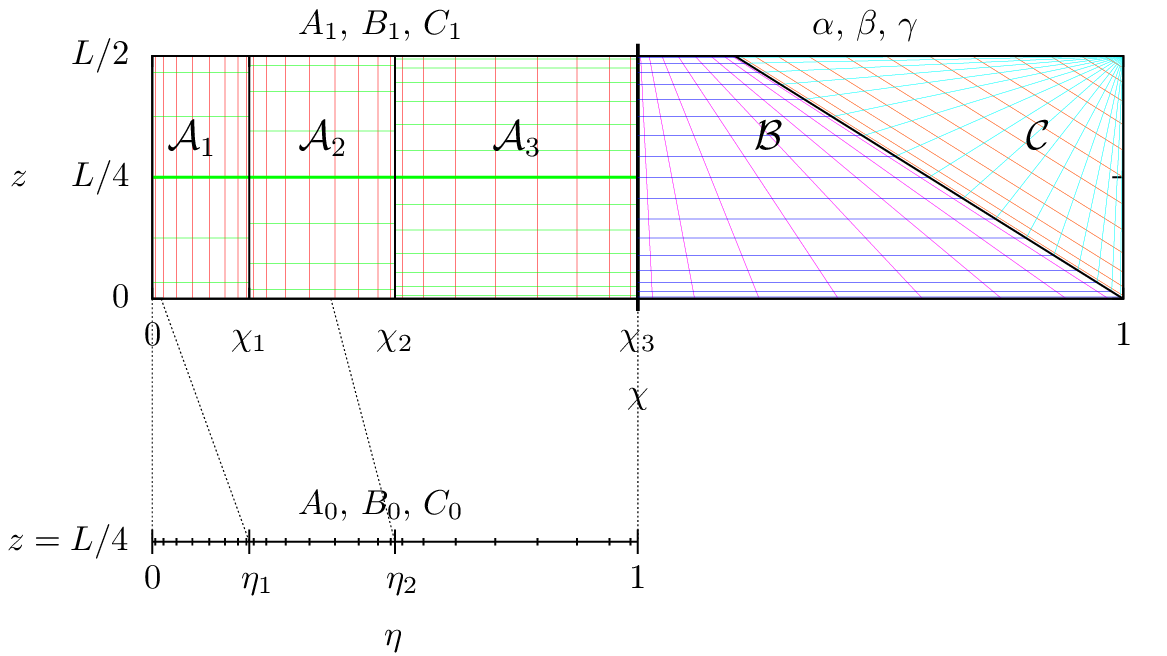}
	\caption{Division of the integration domain into several subdomains for $D=5$. The asymptotic region $\chi\leq\chi_I$ is split up into $I$	rectangles (here $I=3$) with different resolutions regarding 		the $z$-direction. In this region, the numerical algorithm solves for the auxiliary functions 	$A_0$, $B_0$, $C_0$, $A_1$, $B_1$ and $C_1$ which constitute through \eqref{eq:splitAnsatz5D} the metric 	functions $A,B$ and $C$. Note that the one-dimensional functions $A_0$, $B_0$ and $C_0$ are	considered on the $z=L/4$-line (drawn more boldly, see also footnote~\ref{note:oddN}) at gridpoints with 		specific values of the coordinate $\eta$ (cf.~\eqref{eq:eta}). Also, for $\chi\leq\chi _I$ we express the functions $A_1$, $B_1$ and $C_1$ in terms of $\chi$ and $u$ (cf.~\eqref{eq:chi} 						and~\eqref{eq:u}).	The near-horizon region $\chi\ge\chi_I$ is split up into the trapezoidal subdomain $\mathcal B$ and the triangular subdomain $\mathcal C$ which is divided up further, in order to 		resolve steep gradients at the critical point $(\chi, z ) = (1,L/2)$, see fig.~\ref{fig:int_domain_triangle_0}. In $\mathcal B$ and $\mathcal C$, the numerical algorithm solves for the auxiliary 			functions $\alpha $, $\beta $ and $\gamma$ which are related through~\eqref{eq:albega} to $A,B$ and $C$. }
	\label{fig:int_domain_5D}
\end{figure}

In the near-horizon domain $(\chi ,z)\in [\chi _I,1]\times [0,L/2]$ the behavior of the functions is similar to that in the case $D=6$. Therefore, we again introduce the functions $\alpha$, $\beta$ and $\gamma$ according to~\eqref{eq:albega} and utilize the original coordinate $z$. Also, we perform once more an additional domain decomposition into a trapezoidal subdomain $\mathcal B$ and a triangular subdomain $\mathcal C$. The entire setup is illustrated in fig.~\ref{fig:int_domain_5D}. Again, further splitting of the domain $\mathcal C$ turned out to be necessary, in order to enhance the resolution near the critical point $(\chi ,z) = (1,L/2)$, see fig.~\ref{fig:int_domain_triangle_0} and appendix~\ref{appendix:sec:Resolving_the_critical_point} (in both cases, replace $\xi$ by $\chi$).

\section{Results}
\label{sec:Results}

With the help of the numerical approach described in the previous sections we were able to obtain unprecedented horizon deformations characterized through large values of the parameter $\lambda$. In particular, we were able to reach out to
\beq
  			\lambda\lesssim 340 \quad \mbox{for $D=5$} \, , \qquad
  			\lambda\lesssim 202 \quad \mbox{for $D=6$} \, . 
	\label{eq:lambdaMNS}
\eeq
Even for these large numerical limiting values, high accuracy of the corresponding solutions was restored.\footnote{We note that in the large $\lambda$ regime the calculations were carried out within the C-programming language with ``long double'' precision.} In the set-up, we introduced eight subdomains in $D=5$ (with $I=3$ and $J=4$, see figs.~\ref{fig:int_domain_5D} and \ref{fig:int_domain_triangle_0}) and seven subdomains in $D=6$ (with $I=3$ and $J=3$, see figs.~\ref{fig:int_domain} and \ref{fig:int_domain_triangle_0}). Although the spectral resolutions were of the order $N\approx 50$ (in all directions),\footnote{At least, such high resolutions are required for large $\lambda$ in the vicinity of the horizon.} the computational costs were kept at a moderate extent, thus allowing to establish a specific solution on a single computer within a few minutes.\footnote{The most time consuming part of the calculation is the solution of the linear system inside the Newton-Raphson scheme.}

At this point we want to emphasize that our results show qualitatively the same behavior for both spacetime-dimensions considered, $D=5$ and $D=6$. Below we present our findings obtained in the two cases, but we will not stress this qualitative agreement on every occasion.

\subsection{Geometry}
\label{subsec:Geometry}

As will be depicted in appendix~\ref{appendix:sec:Perturbations_around_the_UBS}, the value of $L_{\text{GL}}$ (the critical length of the compact dimension, where the GL instability occurs) can be calculated from linear perturbations in the vicinity of the UBS. The following values were obtained:
\beq
	\frac{L_{\text{GL}}}{r_0} = \begin{cases}
			7.1712728543704(1)  & \text{for } D=5 \,, \\
			4.9516154200735(1)	& \text{for } D=6 \,.
		\end{cases}
	\label{eq:L_GL}
\eeq

\begin{figure}[!ht]
	\includegraphics[scale=1.0]{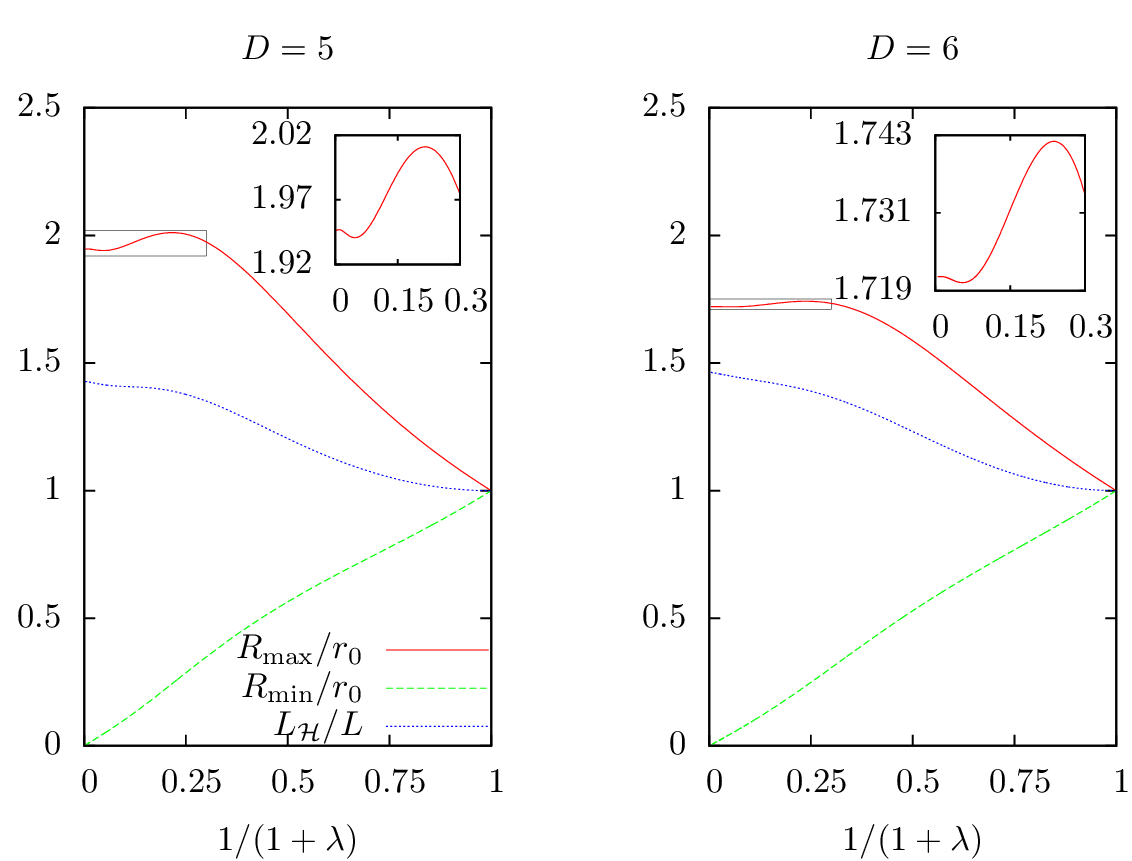} 
	\caption{Maximal and minimal horizon areal radii $\Rmax$ and $\Rmin$ (in units of $r_0$) and the proper length of the compact dimension on the horizon $L_{\mathcal{H}}$ (normalized by $L$) as functions of $1/(1+\lambda )$. The insets (corresponding to the small boxes in the larger picture)  show the region where $\Rmax$ possesses three turning points. Lines of same type correspond to the same quantity in both plots.}
	\label{fig:horizon_quantities}
\end{figure}

In fig.~\ref{fig:horizon_quantities} we show the behavior of representative geometric quantities, defined on the horizon, as functions of $\lambda$. In order to capture the regime of large $\lambda$, we use $1/(1+\lambda )$ as abscissas in the diagrams. In the range considered, the proper length of the compact dimensions $L_{\mathcal{H}}$ (see eq.~\eqref{eq:L_H}) grows monotonically when $\lambda$ is increased while the minimal areal radius $\Rmin$ decreases monotonically and approaches zero for $\lambda\to\infty$ (note that $\Rmax$ remains finite in this limit). Interestingly, $\Rmax$ shows a non-monotonic behavior. In fact, on the way towards our maximally achieved horizon deformation, we encountered three turning points in the $\Rmax$-curve.

\begin{figure}
	\includegraphics[scale=0.99]{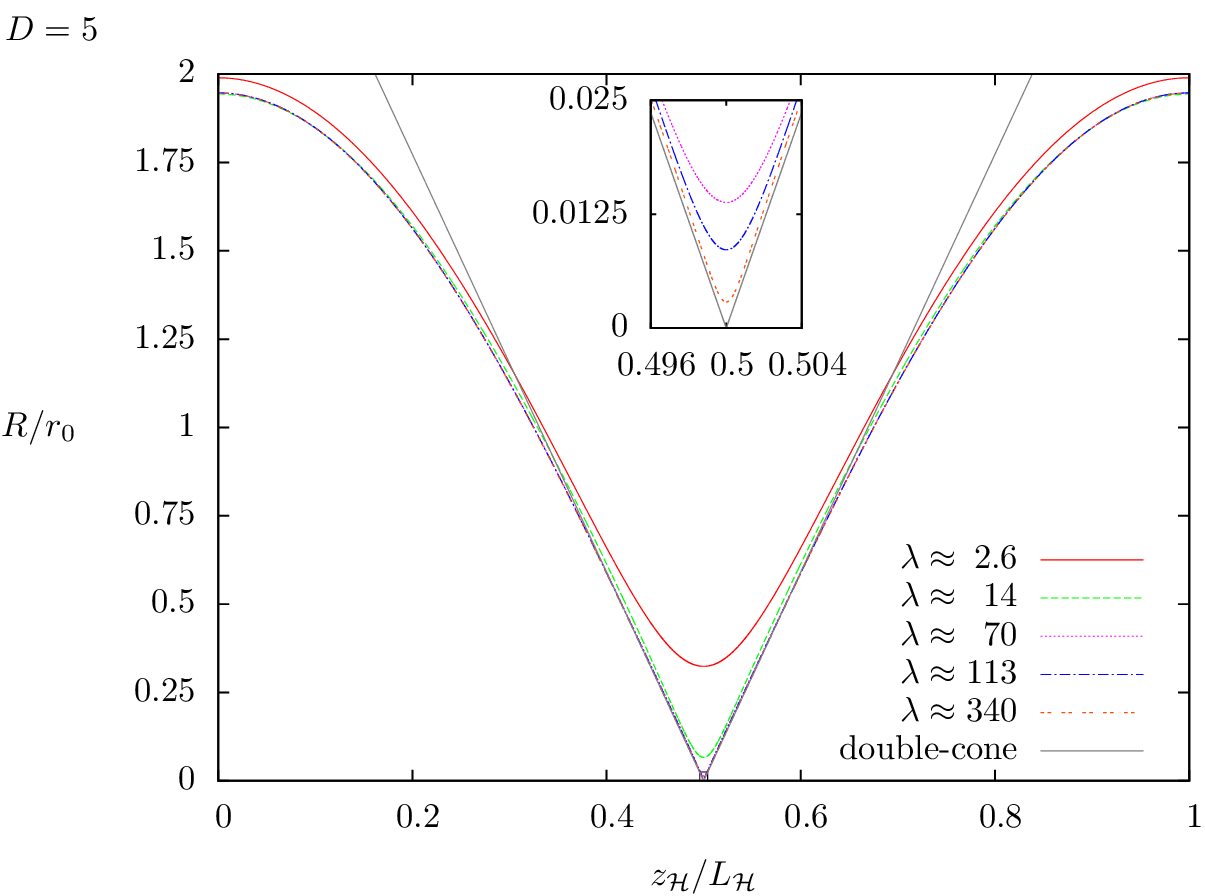}    \\ 	
	\includegraphics[scale=0.99]{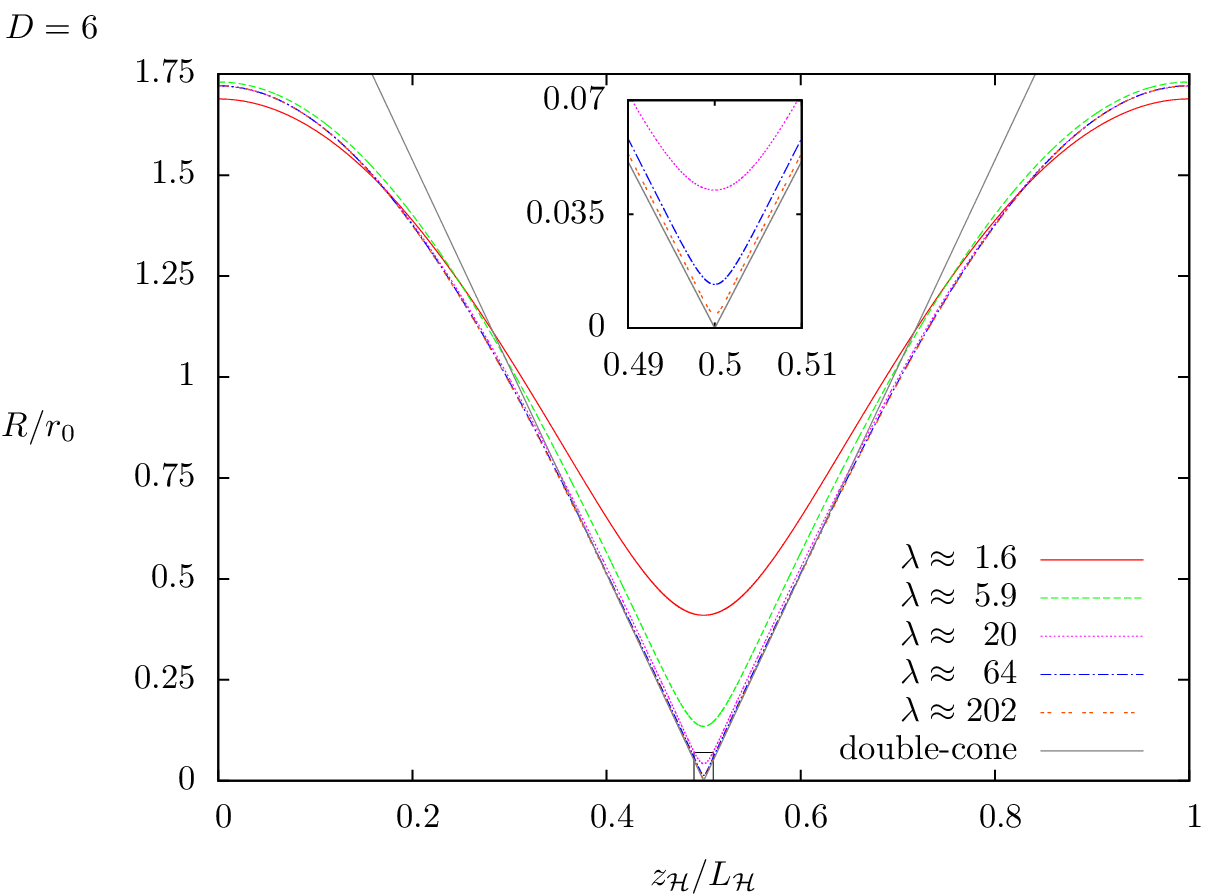}    
	\caption{Horizon areal radius $R$ (in units of $r_0$) for different $\lambda$ along the compact dimension. Proper distances $z_\mathcal{H}(z)=\int_0^z \E ^{B(r_0,\tilde z)} \,\D\tilde z$ are used (cf.~\eqref{eq:L_H}), and the plots are centered around the origin. In the inset, the critical region (marked by a small box in the larger picture) is magnified. Also, the shape of the double-cone geometry is indicated. }
	\label{fig:horizon_1d}
\end{figure}

A more qualitative picture, illustrating the growing deformation of the horizon as $\lambda$ is increased, is given in fig.~\ref{fig:horizon_1d}. Here we plot the horizon areal radius against proper distances in the circle direction. We see a smooth behavior for moderate $\lambda$, while for increasing $\lambda$ the curve pinches off and develops a cusp in the limit $\lambda\to\infty$. 

More concretely, fig.~\ref{fig:horizon_1d} shows that in vicinity of the critical point the horizon approaches the double-cone geometry of~\cite{Kol:2002xz} as $\lambda$ is increased. This supports the conjecture of a phase transition between black strings and localized black holes with the double-cone metric as a local model of the transit solution at the point, where the horizon pinches off~\cite{Kol:2002xz}. 

\subsection{Thermodynamic quantities}
\label{subsec:Thermodynamic_quantities}

\begin{figure}[!ht]
	\includegraphics[scale=1.0]{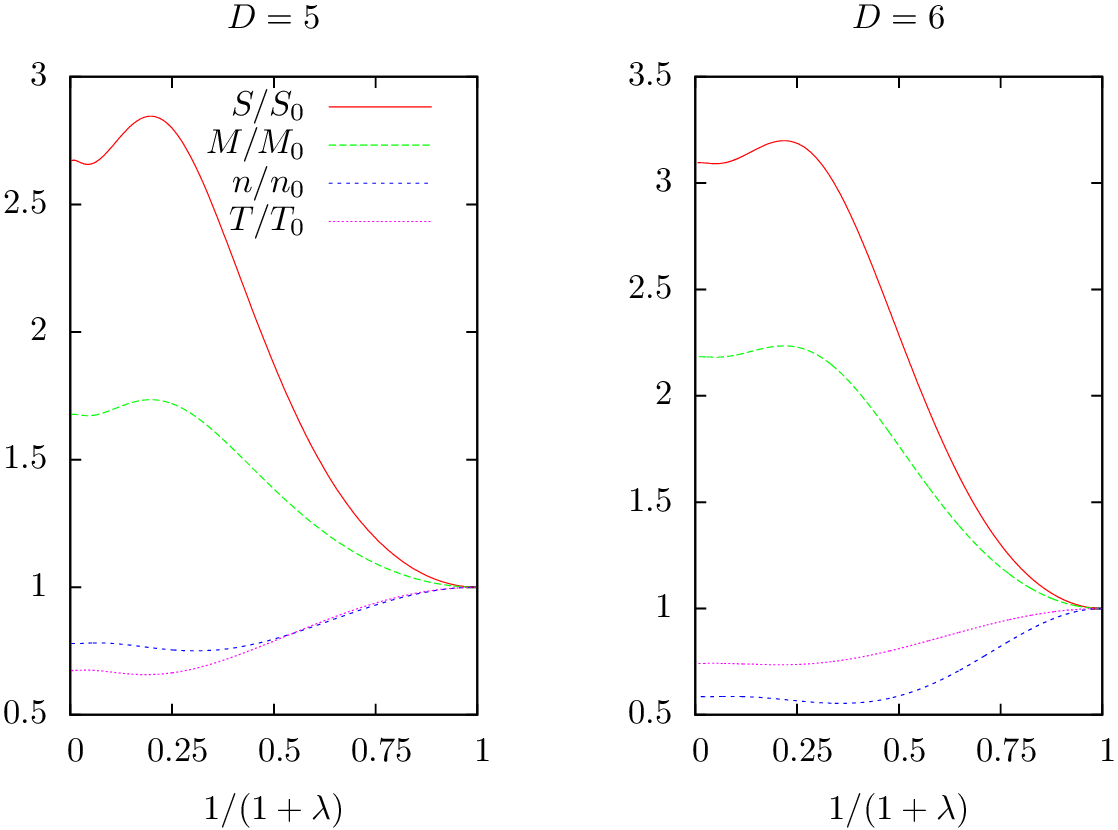} 
	\caption{Entropy $S$, mass $M$, relative tension $n$ and temperature $T$ (normalized by their corresponding UBS values) as functions of $1/(1+ \lambda )$. Lines of same type correspond to the same quantity in both plots.}
	\label{fig:parameter_all}
\end{figure}

\begin{figure}[!ht]
	\includegraphics[scale=1.0]{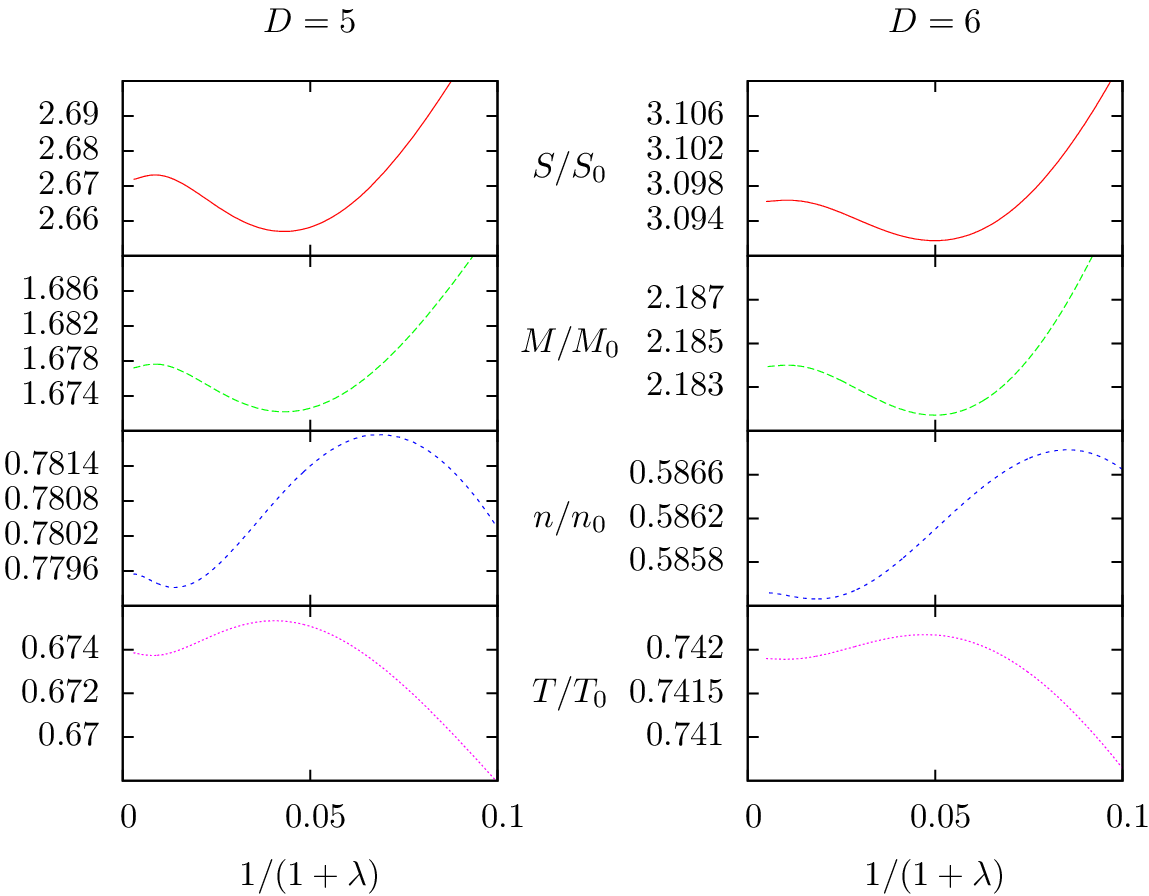}
	\caption{Magnification of the regions in fig.~\ref{fig:parameter_all} where the curves possess two further turning points (apart from their leading turning points).}
	\label{fig:parameter_all_zoom}
\end{figure}

We now turn our focus to representative thermodynamic quantities. In fig.~\ref{fig:parameter_all} we display the behavior of the mass, the relative tension, the temperature and the entropy as functions of $1/(1+\lambda )$. Starting at $\lambda =0$, the mass and the entropy increase with increasing $\lambda$ until a maximum is reached. Similarly, the relative tension and the temperature show a minimum. As $\lambda$ grows further, in each of these functions we observe two additional turning points, appearing on the way to the maximally achieved horizon deformation. These are shown in fig.~\ref{fig:parameter_all_zoom}.

\begin{figure}[!htp]
	\includegraphics[scale=0.99]{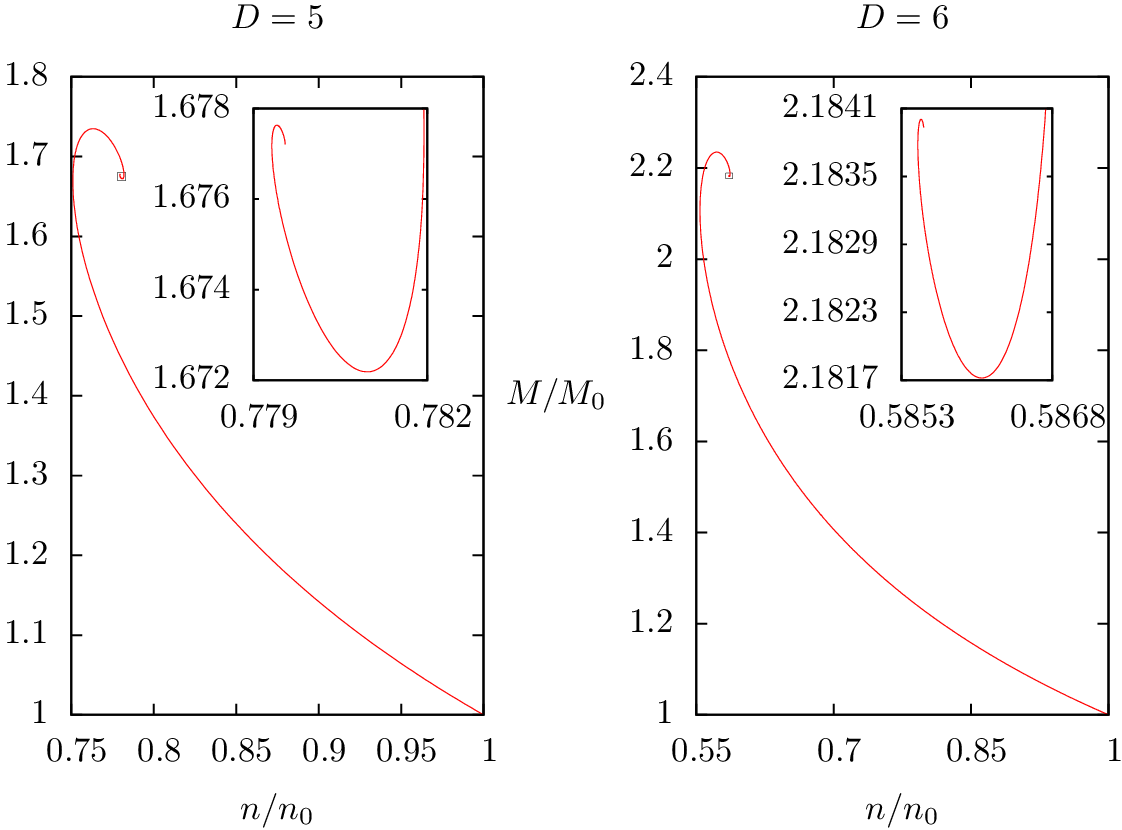}
	\caption{Mass $M$ plotted against the relative tension $n$ (normalized by their corresponding UBS values). The onset of a spiral curve can be seen. The insets show a magnification of the inner part of the spiral. The magnified region is indicated by a tiny box in the larger plot.}
	\label{fig:spiral}
\end{figure}

\begin{figure}[!htp]
	\includegraphics[scale=1.0]{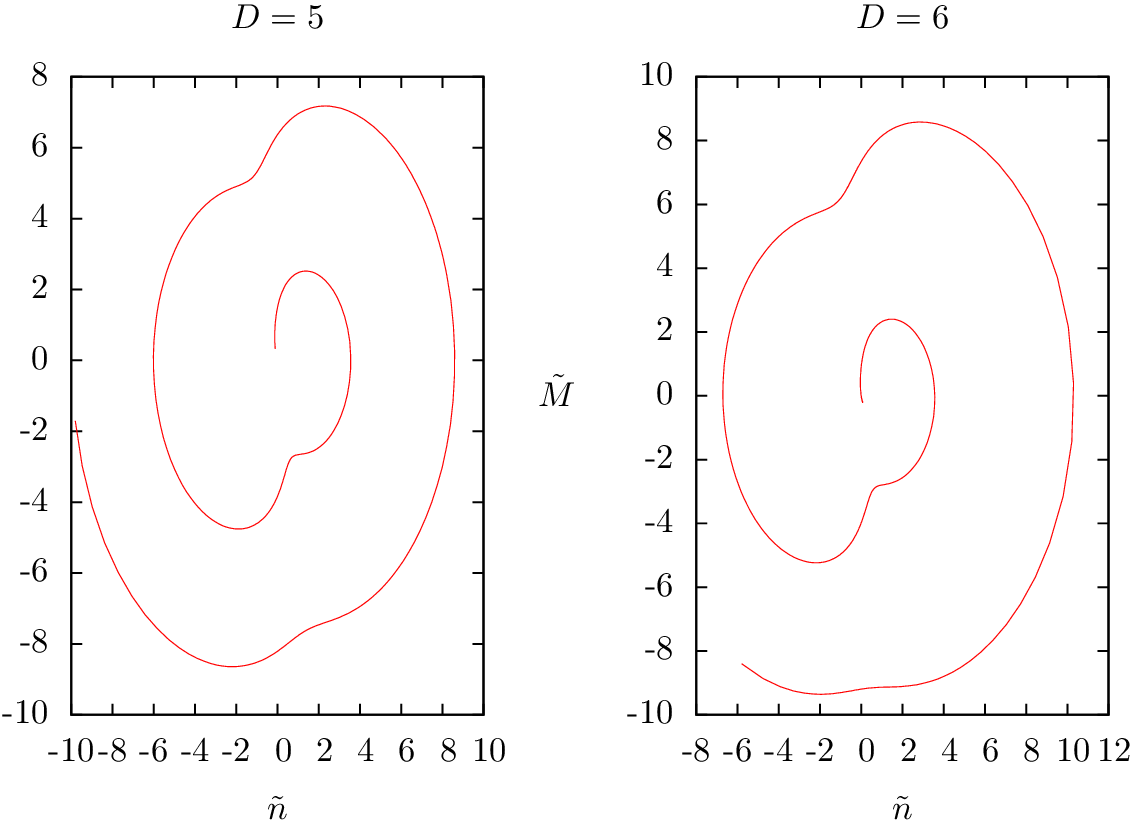} 
	\caption{Logarithmic radial rescaling of fig.~\ref{fig:spiral}. $\tilde M$ and $\tilde n$ are defined in equation~\eqref{eq:logrescale}. } 
	\label{fig:logspiral}
\end{figure}

In~\cite{Harmark:2003dg} it was discussed that the entire black string thermodynamics can be derived from the curve displaying the mass against relative tension. As an immediate consequence of the turning points of $M$ and $n$ along the NBS branch in fig.~\ref{fig:parameter_all} we see a spiral in the phase diagram in fig.~\ref{fig:spiral}. In $D=5$ as well as in $D=6$ the spiral winds about one and half  times (actually, even a bit more) before we reach our maximally numerically attainable value of $\lambda$.
In fig.~\ref{fig:spiral} it becomes apparent that the extent of the spiral twists is shrinking rapidly with each turn. In order to resolve the details we plot the spiral once more in a logarithmically radially rescaled version. To this end,  we choose a point $(n_\text{s},M_\text{s})$ in the phase diagram (fig.~\ref{fig:spiral}) which is located  approximately at the position where the endpoint of the spiral is expected. We center the coordinate system with respect to this point and introduce the radial distance $d = \sqrt{ (n_s-n/n_0)^2 + (M_s-M/M_0)^2}$. Then we define the rescaled quantities
\beq
	\tilde M = \frac{\log d}{d} \left( \frac{M}{M_0} - M_s \right) \quad \text{and} \quad \tilde n = \frac{\log d}{d} \left( \frac{n}{n_0} - n_s \right) \,,
	\label{eq:logrescale}
\eeq
to be plotted against one another in this diagram, see fig.~\ref{fig:logspiral}. Note that this procedure maps inner parts of the spiral to the outside and vice versa and, crucially, it would transform a logarithmic spiral into an Archimedean one. 

Although the spirals in fig.~\ref{fig:logspiral} look rather bumpy, they are well resolved within our logarithmic rescaling. We conclude that in the original phase diagram, fig.~\ref{fig:spiral}, the extent of the spiral twists are shrinking exponentially with each turn, similar to the behavior of a logarithmic spiral.  It is very tempting to conjecture that in the limit $\lambda\to\infty$ the spirals wind up infinitely many times before reaching their endpoints (as it is for a logarithmic spiral). 

Note that spiral curves are likewise exhibited in phase diagrams of other pairs of thermodynamic quantities. However, the $(M,S)$-diagram is an exception. This is due to the first law $\D M = T \D S$, from which we follow that the turning points of mass and entropy coincide. This leads to cusps in the $(M,S)$ diagram rather than smooth twists. 

\subsection{Accuracy}
\label{subsec:Accuracy}

The most interesting features in the figures presented in the previous section concern the regime of large $\lambda$-values and are contained in tiny parts of the diagrams. In order to resolve these details, high accuracy of the solutions is needed such that the uncertainties in the numerically determined values are much smaller than the magnitudes of the small features. In section~\ref{sec:Numerical_implementation} we have described a numerical scheme which is capable to provide such solutions in an adequate amount of time. Here, we want to show that, in the entire regime of $\lambda$-values considered, the results obtained with this scheme are very accurate. For this purpose we consider accuracy tests for the numerical scheme as well as consistency checks given by the physics.

\begin{figure}[!ht]
        \includegraphics[scale=1.0]{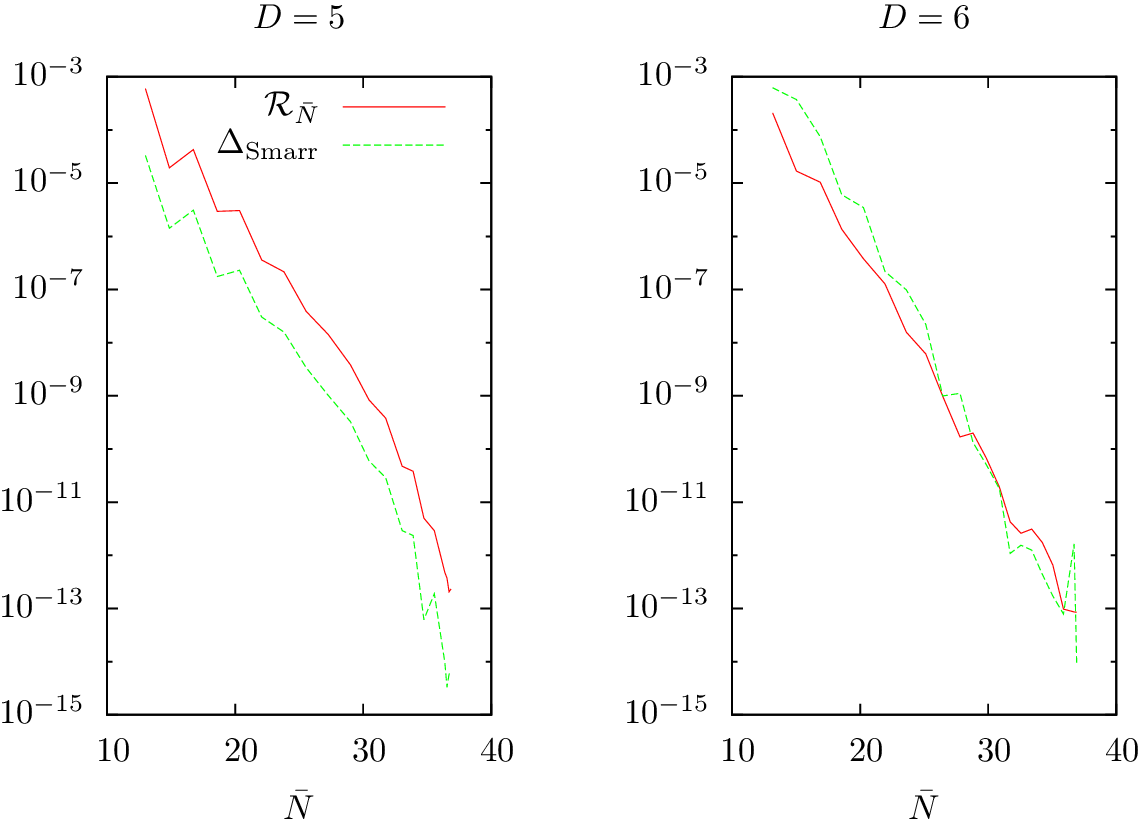}
        \caption{Convergence of the residue $\mathcal R_{\bar N}$ and the deviation from Smarr`s formula $\Delta _{\text{Smarr}}$ for our solutions with largest horizon deformation obtained. Here, the resolution $\bar N$ denotes the mean value, where we have averaged over all subdomains and all directions. Lines of same type correspond to the same quantity in both plots.}
        \label{fig:convergence}
\end{figure}

A commonly used method to measure the accuracy of a spectral algorithm is to compare a reference solution with high resolution with solutions of lower spectral resolution. The comparison is carried out by determining all solutions' function values on a fine  equidistant grid, using spectral interpolation techniques, and then calculating the differences of reference solution to the solutions of lower resolution at each of these grid points. We call the largest magnitude of these  differences the residue $\mathcal R_N$, where $N$ indicates the resolution of the less resolved solution. As $N$ approaches the resolution of the reference solution, $\mathcal R_N$ usually decreases and eventually saturates at a small value due to numerical limitations caused by finite machine precision and rounding errors. In fig.~\ref{fig:convergence} we display the convergence of the residue for the solutions with largest horizon deformation obtained. It can be seen that the residue saturates at values of the order $10^{-13}$, thus illustrating the over-all accuracy of our solutions in the critical regime of large $\lambda$.

Apart from $\mathcal R_N$, figure~\ref{fig:convergence} also shows the difference $\Delta _{\text{Smarr}}$ between right and left hand side of Smarr's formula~\eqref{eq:Smarr}. It converges similarly to the residue, thus Smarr's relation is satisfied very accurately. In addition, another verification of our solutions can be realized by checking the first law of black hole thermodynamics (see eq.~\eqref{eq:1stlaw}). Parametrized with our control parameter $\beta _\text{c}$, the first law reads
\beq
\frac{\D M}{\D \beta _\text{c}} = T \frac{\D S}{\D \beta _\text{c}} \,.
        \label{eq:1stLaw_beta}
\eeq
An accurate way to test eq.~\eqref{eq:1stLaw_beta} on a part of the NBS branch $\beta _{\text{c}_1} \leq \beta _\text{c} \leq \beta _{\text{c}_2}$ is to approximate $M(\beta _\text{c})$, $T(\beta _\text{c})$ and $S(\beta _\text{c})$ by means of a Chebyshev expansion~\eqref{eq:spectral_expansion}, where a solution to the field equations at each Lobatto grid point~\eqref{eq:Lobatto} is needed. The derivatives in~\eqref{eq:1stLaw_beta} can then be performed using standard spectral techniques. As one increases the number of Lobatto grid points, the deviation from the first law becomes smaller. Accordingly, one can plot a typical convergence plot, similar to fig.~\ref{fig:convergence}. In this manner we confirmed that our solutions satisfy the first law on different parts of the NBS branch. The deviation always drops down at least to orders of $10^{-10}$.

It was shown in~\cite{Kudoh:2003ki} that a violation of the constraint equations would not cause any deviation of Smarr's formula or the first law, as long as only the equations~\eqref{eq:field_eqns} are satisfied. Recall that our numerical approach relies on solving~\eqref{eq:field_eqns}, while the constraint equations are only indirectly satisfied by choosing the ``right'' boundary conditions~\cite{Wiseman:2002zc}. Therefore, an examination of the constraint equations by substituting the numerical solution is essential to check if it actually solves {\em all} Einstein's equations. The constraint violations for our solutions with maximal horizon deformations are of the order $10^{-8}$ at the point, where the horizon is supposed to pinch off in the $\lambda\to\infty$ limit. Far from this point the violations are some orders of magnitude smaller, as they are for solutions with moderate horizon deformations.

\section{Summary and Conclusions}
\label{sec:Conclusions}

For the purpose of investigating the critical regime of large $\lambda$-values, corresponding to strong horizon deformations of the NBS, we developed a well-adapted numerical scheme. Using a pseudo-spectral approach we adjusted and extended the methods in a sophisticated manner, in order to get highly accurate solutions in particular in the critical regime. We now discuss the crucial adaptions:

\begin{itemize}
        \item The split of the functions near infinity: \\ In the asymptotic region ($r\to\infty$) we separate each metric function into one part which only depends on $r$ and one which also depends on $z$ (see~\eqref{eq:splitAnsatz6D} and~\eqref{eq:splitAnsatz5D}). The reason for this decomposition is given by the fact that the $z$-dependent modes decay rapidly as $r\to\infty$. From the $z$-independent modes the asymptotic charges can be extracted very accurately, and we obtain higher precision of the overall solution.
This procedure can always be useful in problems where a compactified dimension is present and the modes corresponding to the compact coordinate show a rapid decay at infinity.
        \item Multi-domain splitting and different resolutions in the two directions:\\ We divide our integration domain in a sequence of straightly connected subdomains, see fig.~\ref{fig:int_domain}. Not only in the direction longitudinal to the sequence but also in the transversal direction we use different resolutions, since the rapid decay of the $z$-dependent modes require, for globally highly accurate solutions, a substantially smaller resolution in the asymptotic regime. This leads to non-matching grid points at the common boundary of two neighboring domains.
        \item Coordinate transformations and multiple domains near the horizon: \\ The density of numerical grid points near the critical point is tremendously increased by specific coordinate transformations (see appendix~\ref{appendix:sec:Resolving_the_critical_point}). This enables us to resolve this regime with moderate resolutions very accurately. The corresponding coordinate transformations can be advantageous when strongly pronounced peaks appear in the solutions.
        \item Exponential coordinate transformation: \\ With the help of a coordinate transformation (eq.~\eqref{eq:eta}) it is possible to avoid logarithmic behavior. As a consequence, the spectral convergence rate of the $z$-independent functions in $D=5$ is changed from algebraic to sub-geometric, which significantly decreases the number of required grid points. This coordinate transformation could be useful in many situations, since logarithmic behavior is a frequently emerging problem in numerics.
\end{itemize}

In this work we constructed solutions in $D=5$ and $D=6$ spacetime dimension. We expect that the scheme can easily be generalized to higher dimensions. As logarithmic terms are absent for $D\ge6$, a modification of the asymptotic ansatz~\eqref{eq:splitAnsatz6D} to be incorporated into the routines for $D=6$ should provide the corresponding numerical means.

The most interesting physical result is the appearance of a spiral curve in the black string's phase diagram (see figs.~\ref{fig:spiral} and~\ref{fig:logspiral}) in both the cases $D=5$ and $D=6$. Note that this was already conjectured in~\cite{Kleihaus:2006ee}. In both spacetime dimensions the spiral winds about one and half  times before we reach our maximally numerically attainable deformation of the black string horizon. We conjecture, however, that the spiral winds up infinitely many times in the critical limit $\lambda\to\infty$. Such a behavior was observed in the context of hairy black holes in $AdS_5\times S^5$~\cite{Bhattacharyya:2010yg}. There are further examples of higher dimensional black objects where the beginning of a spiral curve in their phase diagram could be shown, for instance in the context of hairy black holes in global $AdS_5$~\cite{Dias:2011tj} or lumpy black holes~\cite{Dias:2014cia,Emparan:2014pra}. Note that all of these solutions have in common that their branch emanates from the zero-mode of an instability. In the case of the hairy black holes mentioned above it is the superradiant instability of Reisner-Nordstr\" om black holes, while the ultraspinning instability of Myers-Perry black holes leads to the lumpy black holes. Recall that the NBS branch emanates from the GL-instability of the UBS branch. Hence the formation of a spiral curve in the phase diagram seems to be a quite generic feature for such situations.\footnote{We thank Oscar J. C. Dias for pointing this out.}

We conclude by considering some implications of the inspiral of the NBS branch. The conjecture of the black hole/black string phase transition requires that these two branches meet at $\lambda\to\infty$ (from the NBS point of view). It seems to be very likely that the localized black hole branch will likewise pass through a spiral curve which joins to the NBS spiral with a common end point. So far, one turning point in the phase diagram of the localized black holes has been found~\cite{Headrick:2009pv}. Nevertheless, further extension of the localized black hole branch is needed to see evidence for a possible spiral, in particular showing a continuous transition to the NBS spiral. Furthermore, as the turning point in the localized black hole branch goes along with the emergence of an unstable mode~\cite{Headrick:2009pv}, we expect a similar property to be seen in the NBS phase diagram. In other words, if there are infinitely many twists before the two branches meet, then a growing number of unstable modes emerges as one approaches the phase transition.

\section*{Acknowledgments}
We thank Burkhard Kleihaus, Jutta Kunz and Eugen Radu for drawing our attention to this problem and for numerous fruitful discussions. In particular, we thank Burkhard Kleihaus and Jutta Kunz for careful reading of the manuscript. Furthermore, we are grateful to Barak Kol for valuable discussions. This work was supported by the Deutsche Forschungsgemeinschaft (DFG) graduate school GRK 1523/2.

\begin{appendix}
\section{Perturbations around the UBS}
\label{appendix:sec:Perturbations_around_the_UBS}

Perturbation theory in the vicinity of the UBS was first developed in~\cite{Gubser:2001ac} for five dimensions and later applied to six dimensions in~\cite{Wiseman:2002zc}. A generalization to arbitrary dimensions was carried out in~\cite{Sorkin:2004qq}. We want to illustrate, in which manner the study of linear perturbations proves  useful for the development of an accurate and efficient numerical scheme to solve the full set of non-linear equations~\eqref{eq:field_eqns}. Furthermore we derive highly accurate values for $L_{\mathrm{GL}}/r_0$ from this analysis.

For first order perturbations only the marginal GL mode appears and we can write
\refstepcounter{equation} \label{eq:LinPerturb}
\begin{align}
	A &= \varepsilon \, a(r) \cos \left( \tfrac{2\pi}{L}z \right) \, ,
		\tag{\theequation a}  \label{eq:LinPerturbA} \\
	B &= \varepsilon \, b(r) \cos \left( \tfrac{2\pi}{L}z \right) \, , 
		\tag{\theequation b}  \label{eq:LinPerturbB} \\
	C &= \varepsilon \, c(r) \cos \left( \tfrac{2\pi}{L}z \right) \, ,
		\tag{\theequation c}  \label{eq:LinPerturbC}
\end{align}
for some small $\varepsilon$. After substituting this into the field equations, we solely take linear orders of $\varepsilon$ into account. The equations~\eqref{eq:field_eqns} then yield the following set of ordinary differential equations (see also~\cite{Kol:2004pn}):
\refstepcounter{equation} \label{eq:LinPerturbEqns}
\begin{align}
	0 =& f a'' + \left[ \dfrac{3}{2}  f'  + (D-3) \frac{f}{r}  \right] a' + \frac{1}{2} (D-3) f'  c' - \frac{4\pi ^2}{L^2} a \, ,
		\tag{\theequation a}  \label{eq:LinPerturbEqnA} \\
	0 =& f b'' - (D-3) \frac{f}{r} a' + \frac{1}{2} f' b' - \frac{1}{2} (D-3) \left[  f' + 2 (D-4) \frac{f}{r} \right] c' \nonumber \\ 
	   & +  (D-3)(D-4) \frac{b-c}{r^2} - \frac{4\pi ^2}{L^2} b   \, , 
		\tag{\theequation b}  \label{eq:LinPerturbEqnB} \\
	0 =& f c'' + \frac{f}{r} a' + \left[ f' + 2 (D-3) \frac{f}{r} \right] c' - 2 (D-4) \frac{b-c}{r^2} -   \frac{4\pi ^2}{L^2} c \, .
		\tag{\theequation c}  \label{eq:LinPerturbEqnC}
\end{align}

\subsection{6D perturbations}
\label{appendix:subsec:6D_perturbations}

An asymptotic analysis of~\eqref{eq:LinPerturbEqns} for $D=6$ gives rise to the following ansatz:
\refstepcounter{equation} \label{eq:LinPerturbAnsatz6D}
\begin{align}
	a &= \tilde a(r) \,\E ^{-2\pi r/L} \left( \frac{r_0}{r} \right) ^{3/2} \, ,
		\tag{\theequation a} \label{eq:LinPerturbAnsatz6DA} \\
	b &= \tilde b(r) \,\E ^{-2\pi r/L} \, , 
		\tag{\theequation b} \label{eq:LinPerturbAnsatz6DB} \\
	c &= \tilde c(r) \,\E ^{-2\pi r/L} \left( \frac{r_0}{r} \right) \, .
		\tag{\theequation c} \label{eq:LinPerturbAnsatz6DC} 
\end{align}
For the reasons explained in section~\ref{subsubsec:Treatment_of_the_asymptotics_6D}, the introduction of the coordinate $\xi\in [0,1]$ is useful:
\begin{equation*}
	\frac{r_0}{r} = \left[ 1 - (1-\xi )^2 \right] ^2 = \xi ^2 (2-\xi )^2 \, .
	\tag{\ref{eq:xi6D} revisited}
\end{equation*}
It is now straightforward to solve~\eqref{eq:LinPerturbEqns} for $\tilde a$, $\tilde b$ and $\tilde c$ in terms of $\xi$ numerically. This system is a set of {\em homogeneous} linear ordinary differential equations with non-trivial solutions. For a unique solution we need to impose a scaling condition, and we decided to choose $\tilde{c}=1$ at the horizon, i.e.~at $\xi =1$. Analyzing the perturbation equations at the boundaries, we find that at infinity ($\xi =0$), the equations degenerate and the resulting conditions correspond to~\eqref{eq:asympBC10}. On the horizon ($\xi =1$), the equations yield only the two regularity requirements $\tilde a_{,\xi} = 0$ and $\tilde c_{,\xi} = 0$ (cf.~\eqref{eq:BC_horizon_xi}). 

Furthermore, one has to prescribe a value for the dimensionless quantity $L/r_0$ appearing in the equations. Interestingly, the additional regularity requirement $\tilde b_{,\xi}=0$ to be imposed on the horizon is only satisfied for one specific value of $L/r_0$. We compute this value with high numerical precision by considering $L/r_0$ as further unknown  and imposing $\tilde b_{,\xi}=0$ as an additional equation in our numerical scheme. This provides us with the value of $L_{\mathrm{GL}}/r_0$ specified in~\eqref{eq:L_GL}.

Note that the decay of the spectral coefficients of $\tilde a$, $\tilde b$ and $\tilde c$, expressed in terms of the coordinate $\xi$, is geometric, while that of $a$, $b$ and $c$ only is sub-geometric. This is due to the exponential factor $\E ^{-2\pi r/L}$ which is non-analytic at infinity when expressed in terms of $\xi$. 

Apart from an accurate value for $L_{\mathrm{GL}}/r_0$ the solution obtained with this procedure yields an initial guess for the solution of the non-linear equations~\eqref{eq:field_eqns} (which is an essential ingredient for the application of the Newton-Raphson scheme). However, we need to take care of some subtleties in the non-linear regime, which we want to describe now.

As we have seen above, the linear perturbation in the vicinity of the UBS only yields the marginal GL mode (see \eqref{eq:LinPerturb}), i.e. the linear perturbation does not take the asymptotics~\eqref{eq:asymptotics} into account which is carried by the $z$-independent modes. It first occurs in the second order perturbation (see for example~\cite{Wiseman:2002zc}). This leads us to an ansatz in which we split each of the functions $A$, $B$ and $C$ into two parts. One part is $z$-independent and its leading behavior at infinity is given by the asymptotics~\eqref{eq:asymptotics}. Therefore the physical quantities which are encoded in the far field (mass and relative tension, cf.~\eqref{eq:mass} and~\eqref{eq:tension}) follow from the $z$-independent part alone. The remaining part is $z$-dependent and its leading asymptotic behavior corresponds to that of $a$, $b$ and $c$. As explained above, we can get an enhancement of the decay of the spectral coefficients if we extract this behavior analogous to~\eqref{eq:LinPerturbAnsatz6D}, although in the non-linear regime this yields a mere subgeometric convergence, since the exponential factor appears in higher orders. 

The entirety of these considerations leads us to the ansatz~\eqref{eq:splitAnsatz6D}. Note that a sufficient initial guess close to the UBS is obtained by simply neglecting the $z$-independent functions and taking the solutions $\tilde a$, $\tilde b$ and $\tilde c$ of the first order perturbation equations as seed for the $z$-dependent functions.

\subsection{5D perturbations}
\label{appendix:subsec:5D_perturbations}

In the case $D=5$ the asymptotic analysis of~\eqref{eq:LinPerturbEqns} yields the ansatz
\refstepcounter{equation} \label{eq:LinPerturbAnsatz5D}
\begin{align}
	a &= \tilde a(r) \,\E ^{-2\pi r/L} \left( \frac{r_0}{r} \right) ^{1+\pi r_0/L} \, , 
		\tag{\theequation a} \label{eq:LinPerturbAnsatz5DA} \\ 
	b &= \tilde b(r) \,\E ^{-2\pi r/L} \left( \frac{r_0}{r} \right) ^{  \pi r_0/L} \, ,
		\tag{\theequation b} \label{eq:LinPerturbAnsatz5DB} \\
	c &= \tilde c(r) \,\E ^{-2\pi r/L} \left( \frac{r_0}{r} \right) ^{1+\pi r_0/L} \, . 
		\tag{\theequation c} \label{eq:LinPerturbAnsatz5DC}
\end{align}
The introduction of the coordinate $\chi\in [0,1]$ via
\beq
	\frac{r_0}{r} = 1-(1-\chi )^2 = \chi (2-\chi ) \, ,
	\tag{\ref{eq:chi} revisited}
\eeq
means that for our metric functions, which are regular at the horizon, the derivatives with respect to $\chi$ vanish at $\chi =1$. We can therefore solve the corresponding linear ordinary differential equations for $\tilde a$, $\tilde b$ and $\tilde c$ numerically in the same manner as explained in the previous section for $D=6$. In particular, with the scaling condition $\tilde{c}=1$ and the  regularity requirement $\tilde{b}_{,\chi}=0$, both imposed at the horizon, the equations are solved together with corresponding boundary conditions. As for $D=6$, we get simultaneously an accurate value for $L_{\text{GL}}/r_0$ in $D=5$, as specified in~\eqref{eq:L_GL}.

Now, it seems promising to proceed in a similar manner as we did for $D=6$. Indeed, it is a good idea to split the functions $A$, $B$ and $C$ into a $z$-independent and a $z$-dependent part in the non-linear regime. But in contrast to $D=6$ logarithmic behavior in the asymptotics (cf.~\eqref{eq:asymptoticsC}) appears in higher order perturbations.

Let us first concentrate on the $z$-dependent modes. According to~\eqref{eq:LinPerturbAnsatz5D} these modes always carry the exponential factor $\E ^{-2\pi r/L}$. Now, this factor suppresses any logarithmic behavior of $r$ at infinity. For this reason we express the $z$-dependent modes in terms of $\chi$ and refrain from extracting the exponential factor. This yields a sub-geometric convergence of the spectral coefficients of the $z$-dependent modes with respect to $\chi$.

In order to deal with the logarithmic behavior of the $z$-independent modes we make use of the exponential coordinate transformation $\chi =\chi (\eta )$ defined in~\eqref{eq:eta}. As explained in section~\ref{subsubsec:Treatment_of_the_asymptotics_5D}
such terms are now rapidly decreasing with respect to $\eta$ for $\eta\to 0$ and their spectral coefficients converge sub-geometrically. Note that it is essential to extract the leading asymptotics from the $z$-independent modes in order to get the values of $A_\infty$, $B_\infty$ and $C_\infty$.

With this considerations we are led to the ansatz:
\refstepcounter{equation} \label{eq:splitAnsatz5Dbad}
\begin{align}	
	A &=  \hphantom{-} A_0(r) \, \dfrac{r_0}{r} \hphantom{\log \dfrac{r_0}{r}} + \, \tilde A_1(r ,z) \, \cos\left( \tfrac{2\pi}{L}z\right)  \, ,  
		\tag{\theequation a} \label{eq:splitAnsatz5DbadA} \\
	B &=  \hphantom{-} B_0(r) \, \dfrac{r_0}{r} \hphantom{\log \dfrac{r_0}{r}} + \, \tilde B_1(r ,z) \, \cos\left( \tfrac{2\pi}{L}z\right)  \, ,							         
		\tag{\theequation b} \label{eq:splitAnsatz5DbadB} \\ 
	C &=            -  C_0(r) \, \dfrac{r_0}{r}            \log \dfrac{r_0}{r} + \, \tilde C_1(r ,z) \, \cos\left( \tfrac{2\pi}{L}z\right)  \, , 					         
		\tag{\theequation c} \label{eq:splitAnsatz5DbadC} 			  
\end{align}
where we regard the $z$-independent functions $X_0=\{ A_0,B_0,C_0\}$ as functions of $\eta$ and the $z$-dependent functions $\tilde X_1 = \{ \tilde A_1, \tilde B_1, \tilde C_1\}$ as functions of $\chi$ and $z$. 

Now, the equations for $X_0$ are obtained by substituting~\eqref{eq:splitAnsatz5Dbad} into the field eqs.~\eqref{eq:field_eqns} and taking the specific coordinate value $z=L/4$. After cancellation of some prefactors in the resulting equations, their structure on the $\eta$-grid reads as follows:
\beq
	0 = F_0( X_0; \eta) + \left( \dfrac{r(\eta )}{r_0} \right) ^4F_1( \tilde X_1 ; \eta) \, .
	\label{eq:splitEqns5D}
\eeq
Here, $F_0$ depends on the functions $X_0$ and their first and second derivatives with respect to $\eta$, and $F_1$ depends on the functions $\tilde X_1$ and their first derivatives with respect to $z$. Observe the factor $(r/r_0)^4$, which strongly blows up for small values of $\eta$ due to the exponential mapping~\eqref{eq:eta}. Mathematically there is no problem, since the functions $\tilde X_1$ carry the exponential factor $\E ^{-2\pi r/L}$. Hence, for small $\eta$ this suppresses the $(r/r_0)^4$ behavior and eq.~\eqref{eq:splitEqns5D} is dominated by $F_0$. However, {\em numerically} the evaluation of~\eqref{eq:splitEqns5D} for small $\eta$ is highly problematic, since due to finite machine precision, one can not guarantee that the functions $\tilde X_1$ are always small enough to compensate the $(r/r_0)^4$ behavior at each stage of the Newton-Raphson scheme. We tackle this technical problem by  rewriting the functions $\tilde X_1$ as
\beq
	\tilde X_1 = \left( \dfrac{r_0}{r} \right) ^4 X_1 \, ,
	\label{eq:split2DFunctions5D}
\eeq
which means that the problematic term $(r/r_0)^4$ in~\eqref{eq:splitEqns5D} cancels. Finally, by expressing $(A,B,C)$ in terms of $X_1 = \{ A_1, B_1, C_1 \}$ (as well as $X_0$), we obtain the ansatz~\eqref{eq:splitAnsatz5D}. We note that the fall-off of the spectral coefficients of the functions $X_1$ with respect to $\chi$ is slower than that of $\tilde X_1$ but it still is sub-geometric and we thus get high accuracy with reasonable resolutions.

For constructing an initial guess for the Newton-Raphson scheme in the vicinity of the UBS, we proceed similarly as in $D=6$: We set the $z$-independent functions to zero and obtain the $z$-dependent functions from the solution of the first order perturbation equations.

\section{Resolving the critical point}
\label{appendix:sec:Resolving_the_critical_point}

\begin{figure}[!ht]
	\includegraphics[scale=0.96]{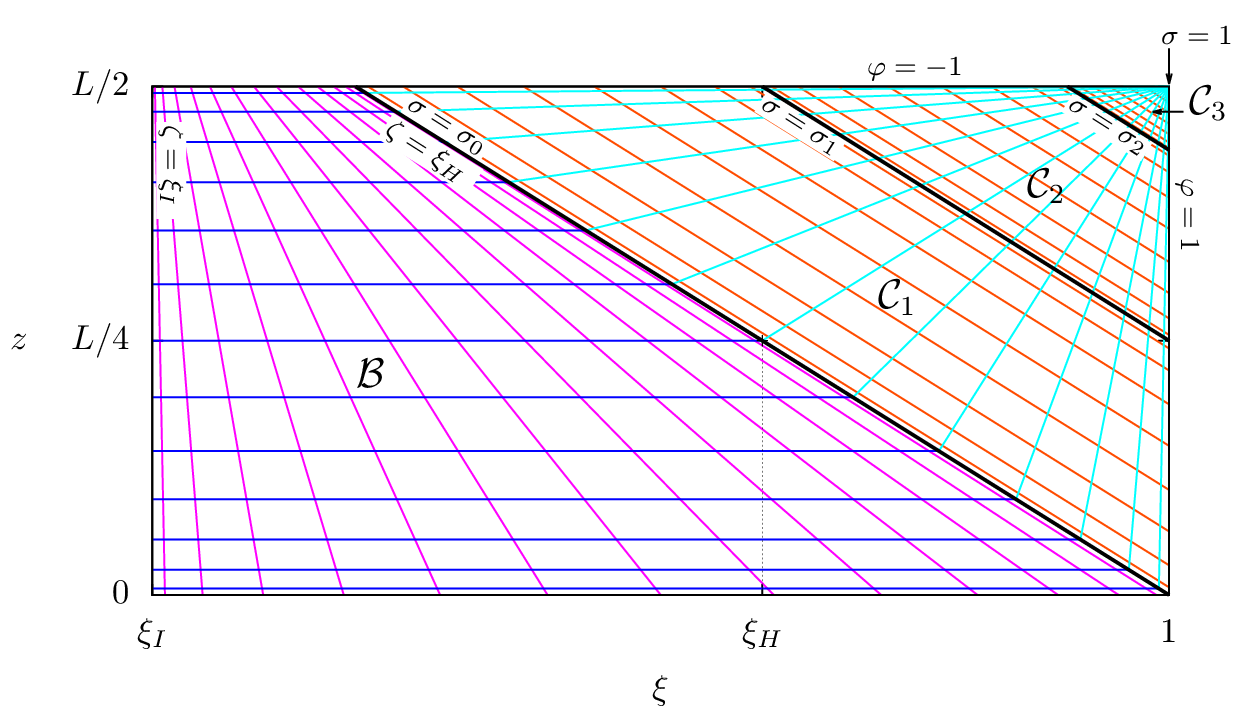}
	\caption{Domain setup in the vicinity of the horizon, i.e.~for $\xi \geq\xi _I$. The triangular domain is subdivided into $J$ layers surrounding the critical point (here $J=3$). The following coordinate lines were drawn: $\zeta = \text{const.}$ (violet), $z = \text{const.}$ (dark blue), $\sigma = \text{const.}$ (orange) and $\varphi = \text{const.}$ (light blue).}
	\label{fig:int_domain_triangle}
\end{figure}

We want to describe the domain splitting and the coordinate transformations which provide us with a high spatial resolution particularly in vicinity of the critical point $(\xi ,z) = (1,L/2)$.\footnote{The consideration in this section is made for the spacetime dimension $D=6$ with the preferred coordinate $\xi$ in the vicinity of the horizon. For $D=5$ one merely has to replace $\xi$ by the coordinate $\chi$ preferred in that case, e.g.~the critical point in $D=5$ is given by $(\chi ,z) = (1,L/2)$.}
A first step 
is the decomposition of the region $\{ (\xi ,z): \xi _I\leq\xi\leq 1 \, , ~ 0\leq z\leq L/2 \} =[\xi _I,1]\times [0,L/2]$ into the trapezoidal domain $\mathcal B$ and the triangular domain $\mathcal C$, see fig.~\ref{fig:int_domain}. In terms of the new coordinate  $\zeta\in [\xi _I ,\xi _H]$  with 
\beq
	\zeta (\xi ,z) = \xi _I + \frac{(\xi _H - \xi _I) (\xi - \xi _I )}{(\xi _H-\xi _I) + (1-\xi _H) ( 1-4\, z/L )} \, ,
	\label{eq:zeta}
\eeq
the domain $\mathcal B$ is mapped onto the rectangle
$\{ (\zeta ,z): \xi _I\leq\zeta\leq\xi_H \, , ~ 0\leq z\leq L/2 \} =[\xi _I,\xi_H]\times [0,L/2]$. In particular, $\zeta =\xi _H$ corresponds to the hypotenuse of the triangular domain, which can be described by the equation $1-4\, z/L = (\xi -\xi _H)/(1-\xi _H)$, see fig.~\ref{fig:int_domain_triangle}.
 
The triangular domain $\mathcal C$ is mapped by means of the following coordinate transformation
\begin{align}
	\sigma  (\xi , z) &= 1 - \frac{1}{2}\left[ (1-\xi ) + (1-\xi _H)(2 -4\, z/L ) \right] \, , \label{eq:sigma} \\
	\varphi (\xi , z) &= 1 - 2\frac{1-\xi}{(1-\xi ) + (1-\xi _H)(2 -4\, z/L)} \,  \label{eq:phi}
\end{align}
onto the  rectangle $\{ (\sigma ,\varphi): 
\xi_H\leq\sigma\leq 1 \, , ~-1\leq \varphi\leq 1 \} =[\xi_H,1]\times [-1,1]$. 
Crucially, coordinate lines 
of constant $\varphi$-values converge towards the critical point, which can be seen from the inverted form of eqs.~\eqref{eq:sigma} and~\eqref{eq:phi}:
\begin{align}	
	\xi         &= 1-(1-\sigma )(1-\varphi ) \, , \label{eq:xi_of_sigma_phi} \\
	1 - 4\, z/L &= \frac{(1-\sigma )(1+\varphi )}{1-\xi _H} -1 \, . \label{eq:u_of_sigma_phi}
\end{align}
For all $\varphi\in[-1,1]$ the coordinate value $\sigma =1$ corresponds to the critical point $(\xi,z)=(1,L/2)$, which means that this single point in the $(\xi ,z)$-chart is blown up to an edge in the $(\sigma ,\varphi )$-chart. The remainder of the horizon ($\xi =1, z<L/2$) corresponds to $\varphi =1,\sigma<1$ and the hypotenuse of the triangular domain is obtained for $\sigma = \xi _H$. Finally, $z = L/2$ is associated with $\varphi = -1$.

The domain decomposition and coordinate mappings described above allow us to use Chebyshev expansions on the several rectangular domains and, moreover, to obtain, in our discretized numerical scheme, densely distributed grid points in the vicinity of the critical point. However, the steep gradients at that point require still a careful treatment which we address in particular through the following two additional steps.

Similar to the decomposition of the domain ${\mathcal A}$ in section~\ref{subsubsec:Decomposition_of_the_numerical_domain_6D}, we subdivide the triangular domain $\mathcal C = \{ (\sigma ,\varphi ): \xi _H\leq\sigma\leq 1 \, , ~ -1\leq\varphi\leq 1 \} = [\xi _H,1]\times [-1,1]$ into $J$ further subdomains $\mathcal C_j =[\sigma _{j-1},\sigma _j]\times [-1,1]$, where $j=1,2,\ldots ,J$ and $\xi _H=\sigma _0 < \sigma _1 < \ldots < \sigma _J =1$. The benefits are similar to those described in the previous context. We illustrate this domain subdivision in fig.~\ref{fig:int_domain_triangle}.

\begin{figure}[!ht]
	\includegraphics[scale=0.96]{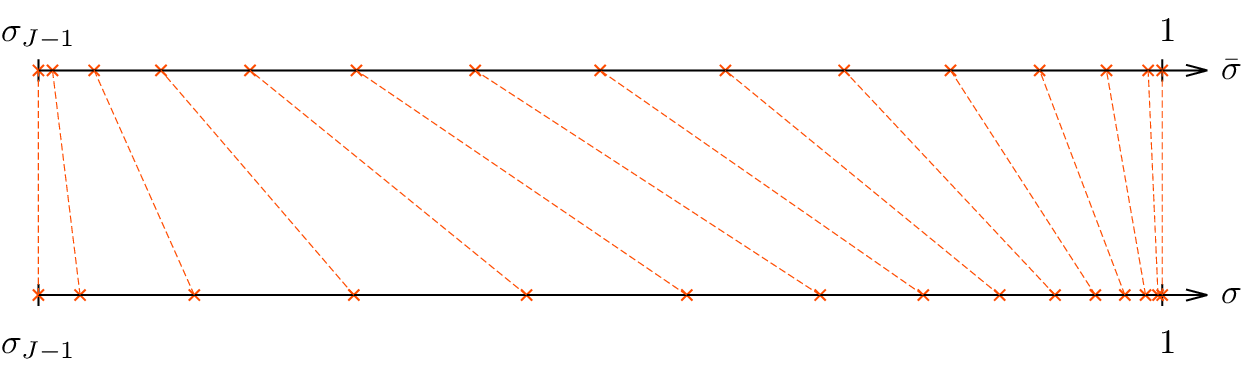}
	\caption{Lobatto grid points~\eqref{eq:Lobatto} on the $\bar\sigma$-grid mapped to the corresponding values on the $\sigma$-grid according to~\eqref{eq:sinh_trafo}. Here we took $N=15$ and $\kappa =3$.}
	\label{fig:sinh_trafo}
\end{figure}

Finally, we resolve steep gradients at the critical point through an analytic mesh-refinement carried out within the triangular subdomain $\mathcal C_J$ which contains the critical point. The analytic mesh-refinement is defined by the mapping
\beq
	\sigma (\bar\sigma ) = 1 - (1-\sigma _{J-1}) \frac{\sinh \left( \kappa \frac{1-\bar\sigma}{1-\sigma _{J-1}} \right)}{\sinh \kappa} \, ,
	\label{eq:sinh_trafo}
\eeq
where the new coordinate $\bar\sigma$ is located in $[\sigma _{J-1},1]$. Depending on the parameter $\kappa$, the gridpoints in the $\bar\sigma$-chart, chosen according to~\eqref{eq:Lobatto}, are densely distributed about $\sigma=1$ in the $\sigma$-chart. On the other hand, the mesh is coarser at the opposite edge, $\sigma=\sigma _{J-1}$, see fig.~\ref{fig:sinh_trafo}. 

This coordinate transformation proved to be appropriate to resolve steep gradients as demonstrated in~\cite{meinel2012relativistic,Macedo:2014bfa} (see~\cite{Ammon:2016szz} for a recent application). The parameter $\kappa >0$ has to be chosen such that the fall-off of the spectral coefficients of the solution is as rapid as possible (there is an optimal $\kappa$, mostly of $\mathcal O(1)$). With this analytic mesh-refinement, the spectral coefficients according to $\bar\sigma$ show a more rapid decay as compared to the fall-off obtained when using $\sigma$. Thus, the costs of the numerical scheme are reduced and, moreover, we observe a substantial increase of the accuracy of the results.

\end{appendix}

\bibliographystyle{JHEP}
\bibliography{NBS}

\end{document}